\documentclass[%
 aip,
 amsmath,amssymb,
 reprint,%
]{revtex4-1}
\usepackage[utf8]{inputenc}
\makeatletter
\usepackage{graphicx}
\usepackage{dcolumn}
\usepackage{subcaption}
\usepackage{caption}
\usepackage{mathtools}
\usepackage{bm}

\usepackage[utf8]{inputenc}
\usepackage[T1]{fontenc}
\usepackage{mathptmx}
\usepackage{etoolbox}
\usepackage{orcidlink}

\DeclarePairedDelimiter\autobracket{\langle}{\rangle}
\newcommand{\br}[1]{\autobracket*{#1}}
\newcommand*{\rom}[1]{\expandafter\@slowromancap\romannumeral #1@}
\makeatletter
\def\@email#1#2{%
 \endgroup
 \patchcmd{\titleblock@produce}
  {\frontmatter@RRAPformat}
  {\frontmatter@RRAPformat{\produce@RRAP{*#1\href{mailto:#2}{#2}}}\frontmatter@RRAPformat}
  {}{}
}%
\makeatother

\hypersetup{
    colorlinks=true,
    linkcolor=blue,
    urlcolor=blue
}
\renewcommand{\abstractname}{Abstract}
\begin{document}

\preprint{AIP/123-QED}

\title[Understanding Plasma Turbulence Through Exact Coherent Spacetime Structures]{Understanding Plasma Turbulence Through Exact Coherent Structures}
\author{Sidney D.V. Williams \orcidlink{0000-0001-9837-5895}}
 \affiliation{Department of Physics, University of California San Diego, La Jolla, CA 92093}
 \affiliation{Center for Energy Research, University of California San Diego, La Jolla, CA 92093}
\author{Matthew N. Gudorf \orcidlink{0009-0005-8506-2052}}%
\affiliation{School of Physics, Georgia Institute of Technology, Atlanta, GA 30332}%

\author{Dmitri M. Orlov \orcidlink{0000-0002-2230-457X}}
\affiliation{Center for Energy Research, University of California San Diego, La Jolla, CA 92093}%

\date{\today}

\begin{abstract}
Plasma turbulence is a key challenge in understanding transport phenomena in magnetically confined plasmas. This work presents a novel approach using periodic orbit theory to analyze plasma turbulence, identifying fundamental structures that underpin chaotic motion. By applying numerical optimization techniques to the Kuramoto-Sivashinsky equation—a reduced model for drift-wave-driven trapped particle turbulence—we extract coherent spacetime patterns that serve as building blocks of turbulent dynamics. These structures provide a framework to systematically describe turbulence as a composition of recurrent solutions, revealing an underlying order within chaotic plasma motion. Our findings suggest that multi-periodic orbit theory can be effectively applied to spatiotemporal turbulence, offering a new method for predicting and potentially controlling transport processes in fusion plasmas. This study provides a bridge between nonlinear dynamical systems theory and plasma physics, highlighting the relevance of periodic orbit approaches for understanding complex plasma behavior.
\end{abstract}

\maketitle

\section{Introduction}
Deterministic chaos is ubiquitous within plasma physics. Applied Resonant Magnetic Perturbations (RMPs) and other 3D fields, such as intrinsic error fields (EFs), lead to chaotic field lines. These lines can then induce kinetic chaos through the charged particles that interact with them\cite{moges_kinetic_2024}. Chaos also manifests in a magnetically confined plasma through turbulence--often treated as the holy grail of chaos theory--which is the driver of most of the transport across magnetic field lines \cite{Horton_Review}. This results in increased, seemingly random, particle transport out of the plasma core. Chaos is not, however, random. There is a distinct structure to the complex motion, and evidence of this can be seen throughout experimental, and theoretical studies. 

In plasma turbulence, structure manifests in many ways. These include phase space granulation, forming "holes" \cite{Dupree1982} and "clumps" \cite{dupree_theory_1972}; nonlinear interactions among Alfvén waves, which generate a broad spectrum of Alfvén modes and quasimodes \cite{Dorfman2024}; and, at the edge of fusion devices, the emergence of coherent "blobs" (density enhancement events) and "holes" (density depletion events) \cite{carter_intermittent_2006}. More recently, relative periodic orbits—defined below—have been identified as key structures in the transition to turbulence in plasma \cite{Smith2025}.

Structure also dominates when considering 3D field effects, where Hamiltonian structure of the magnetic field becomes important. Resonant fields from external magnetic perturbations may penetrate into the plasma core regions and open magnetic islands on rational surfaces. These topological structures (magnetic islands) can create stochastic (really deterministically chaotic) regions by two primary mechanisms. One is through the magnetic island’s hyperbolic X-point and interactions of heteroclinic tangles, which often result in stochastic regions localized near the island X-points\cite{EvansCPP}. These tangle structures, resulting from the splitting of the stable and unstable manifolds when 3D fields are applied, have been experimentally observed in the DIII-D tokamak\cite{TE_Evans_2005} for the primary (axisymmetric) X-point. The second mechanism is the overlap of island chains located on different rational surfaces, often characterized by the Chirikov parameter\cite{Lichtenberg_Lieberman_1983}. This scenario leads to a stochastic edge with remnant magnetic islands working as attractors. 

This all begs the question: can we use this structure, and is it the symptom or the cause?  

Answering this is central to periodic orbit theory, a subfield of chaos theory. Conceptually, the easiest entry point to this subfield is to begin by describing the behavior of chaotic, dissipative systems. Through the dissipation mechanism (viscosity, resistivity, nonlinear eddy-eddy interactions, etc.), the small-scale fluctuations are rapidly damped, leading a chaotic dissipative system to collapse to a finite-dimensional "inertial manifold" \cite{ding_estimating_2016, ashtari_adjoint-based_2024}. Within this chaotic attractor, there exist non-chaotic, but dynamically unstable solutions manifesting as saddles within state space\cite{Devaney_1989}. These solutions include: equilibria; traveling waves--equilibria in a co-moving reference frame; periodic orbits--the system's state repeating after some temporal or spatial period; and relative periodic orbits--a periodic orbit up to a discrete spatial translation. Each of these solutions, in some sense, are invariant under the application of the system's governing equations. Invariance, along with instability, defines a rigid structure of chaotic dynamics through the stable and unstable manifolds of each solution: an arbitrary trajectory is "sucked in" along the stable manifold and "spat out" along the unstable manifold. This picture forms the basis of periodic orbit theory, which formalizes the idea that recurrent structures define a skeleton for chaotic dynamics\cite{budanur_relative_2017}.

In periodic orbit theory, provided that we have collected enough periodic orbits such that their unstable manifolds trace out all neighborhoods of important recurring flows, a hierarchy of orbits is established, and it becomes possible to converge cycle-averaging formulae that provide long-time averages of physical observables\cite{Cvitanović_2013,ChaosBook}. These formulae are additive and rely on weights that are the same for any chosen observable\cite{CBCycleExpandOb}. Unfortunately, with spatially extended systems (say long, turbulent pipe flow), finding the number of periodic orbits needed to trace out all necessary state space becomes computationally intractable with traditional time-integrated dynamical systems methods\cite{budanur_relative_2017}. In attempts to fix this, data-driven techniques\cite{pughe-sanford_computing_2023}, and deep neural networks defining a Markovian view of turbulence\cite{page_recurrent_2024} have been used in lieu of the machinery of periodic orbit theory. In both cases, the strategy boils down to determining the weights that appear numerically in averaging procedures instead of using a closed formula. Nominally, this fixes some of the issues with periodic orbit theory. Pughe-Sanford et al.'s method circumvents the need for vast libraries of unstable periodic orbits, whereas Page et al. found that keeping only the idea of shadowing was sufficient to self-consistently determine orbits and weights through gradient descent and deep learning. 

A more physics-inspired approach is to deal with the wide range of spatiotemporal patterns generated by an extended system by allowing a whole multi-periodic (periodic in both space and time) spacetime pattern to stand where only a time-periodic orbit did in periodic orbit theory\cite{Cvitanovic2025chaotic,liang_chaotic_2022}. In this new picture, turbulence is a guided walk through a catalog of allowed shapes and swirls that constitute the spatiotemporal solutions allowed by the governing equation(s)\cite{cvitanovic_chaotic_2000}. Plasma turbulence is inherently a chaotic spacetime field theory, so it is an ideal physically relevant system to apply this multi-periodic orbit theory to.  

Instead of attempting to treat every minutiae of plasma turbulence, the goal here is to introduce a new conceptual and computational framework in which to approach the chaotic motion ubiquitous in modern fusion devices, without unnecessary complexity. The model and the mode we have chosen reflect this goal. Trapped Ion Modes (TIMs) are a class of trapped particle modes that share much of the physics and linear characteristics of Trapped Electron Modes (TEMs) but are more tractable numerically  \cite{Mandal_2023,GDepret_2000}. Additionally, TIMs provide a prototype for kinetic instabilities due to the resonance condition central to their existence\cite{GDepret_2000}. The chosen model--the Kuramoto Sivashinsky Equation--is often used as a proving ground for new theories within nonlinear dynamics in lieu of the Navier Stokes Equations. This is because it models the fluttering flame front of a ring of fire\cite{sivashinsky_flame_1980,sivashinsky_nonlinear_1977,kuramoto_diffusion-induced_1978}: a "simple" reduced fluid system, capturing the essential physics and mathematical complexity of Navier Stokes while retaining tractability. Similarly, we use the Kuramoto Sivashinsky Equation as a reduced model for drift-wave driven plasma turbulence. 

Although named after its connection to flame fronts, the Kuramoto Sivashinsky equation was originally derived by LaQuey et al. in 1975\cite{laquey_nonlinear_1975} and expanded upon by Cohen et al. a year later\cite{Cohen_1976} to explore the saturation of dissipative TIMs. The essentials of their derivation will be reviewed in Section \ref{second}. In Section \ref{third}, the concepts of periodic and multi-periodic orbit theory needed for a data-driven application to TIMs will be reviewed. In Section \ref{fourth}, the numerical methods used will be elucidated, and the advantages of shifting to a field-theoretic viewpoint will be explored. Finally, Section \ref{fifth} will directly apply multi-periodic orbit theory to the Kuramoto Sivashinksy equation, demonstrating how fundamental blocks of turbulent motion can be used to construct large, chaotic solutions, as well as compute averages of physical observables.        

\section{Trapped Ion Modes and The Kuramoto Sivashinsky Equation} \label{second}
Trapped Ion Modes are conceptually very similar to Trapped Electron Modes, being driven by functionally the same physics, just with different scales. TIMs are drift waves at the ion precession frequency that are driven unstable through strong ion temperature gradients, gradients in ion density, and resonance with ions trapped in "banana orbits". In modern machines, TIMs have been overshadowed by Trapped Electron Modes due to the former's much larger banana width damping out the gyro-bounce average operator, which goes as $1+\frac{k_{\perp}^2\delta^2_{bi}}{4}\sim J_0(\delta_{bi}k_{\perp})$\cite{drouot_gyro-kinetic_2014}. Where $J_0$ is the zeroth-order Bessel function of the first kind.

In the context of TIMs, the Kuramoto Sivashinsky Equation is a reduced model of the dissipative Trapped Ion Mode derived using kinetic corrections to Kadomtsev and Pogutse's slab model continuity equation\cite{Kadomtsev_1971,Kadomtsev_review} closed through the trapped particle quasineutrality condition\cite{laquey_nonlinear_1975,Cohen_1976,Cohen_1978}:
\begin{equation}\label{fluideqns}
    \begin{split}
        \partial_tn^T_{e,i}+\frac{c}{B}(\hat{z}\times\nabla\varphi)\cdot\nabla n_{e,i}^T=-\nu_{\mp}\left(n_{e,i}^T-n_0\epsilon^{1/2}e^{\pm e\varphi/T}\right)\\
        n_{e}^T+n_0(1-\epsilon^{1/2})e^{-e\varphi/T}=n_{i}^T+n_0(1-\epsilon^{1/2})e^{e\varphi/T}+\delta n
    \end{split}
\end{equation}
Here, the RHS of the continuity equation is driving the trapped particle density towards its equilibrium Boltzmann distribution: $\epsilon^{1/2}n_0\exp{(\pm e\varphi/T)}$, and $\delta n$ represents the small kinetic effects neglected in the base fluid model\cite{Cohen_1978}. We will be asserting the appropriate time scales for this mode: $\omega_{Bi}\gg\nu_-\gg\partial_t\sim k_yV_*\gg\nu_+$ where $\omega_{Bi}$ is the ion bounce frequency, $\nu_{\pm}\sim\epsilon^{-1/2}\nu_{i,e}$ are the effective ion-ion and electron-electron collision frequencies respectively, and $V_*$ is the electron diamagnetic velocity.

Following Cohen et al.\cite{Cohen_1978}, \eqref{fluideqns} can be combined to find an equation describing the normalized potential fluctuations, $\phi\equiv e\varphi/T$. This is accomplished by taking the difference of the ion and electron continuity equations, expanding to second order in $\phi$ and first order in $\nu_-^{-1}$, and solving for $n_e^T$: 
\begin{equation}\label{neapprox}
    \begin{split}
        n_{e}^T\approx\epsilon^{1/2}n_0(1+\phi+\phi^2/2)\\
        +\nu_-^{-1}\frac{D}{Dt}[2n_0(1-\epsilon^{1/2})\phi-\delta n]\\
        +\frac{\nu_+}{\nu_-}[2n_0\phi-\delta n]
    \end{split}
\end{equation}
Where
\begin{equation}\label{dt}
\begin{split}
    \frac{D}{Dt}\equiv\partial_t-2\epsilon^{-1/2}V_*(\partial_x(\ln n_0))^{-1}(\partial_x\phi\partial_y-\partial_y\phi\partial_x)\\
    +2\epsilon^{-1/2}V_*\eta_i\phi\partial_y\\
    V_*\equiv V_*^T\equiv-\frac{\epsilon^{1/2}}{2}\frac{cT}{eB}\partial_x\ln (n_0),\hspace{3pt}\eta_i\equiv\frac{\partial\ln T}{\partial\ln n_0}<\frac{2}{3}
\end{split}
\end{equation}
Throughout the rest of the paper we shall use the notation "$V_*$" to denote the trapped particle diamagnetic drift velocity. Note that quasineutrality was used to solve for $n_i^T$ in \eqref{neapprox}. From here, \eqref{neapprox} is plugged into the electron continuity equation. We truncate at $O(\phi^2)$ and $O(\delta n)$, enforce $\epsilon<1$, and utilize the linear first order electron drift wave dispersion relation $\partial_t\phi\approx V_*\partial_y\phi$. This yields
\begin{equation}\label{2DKSE}
\begin{split}
    \left[\partial_t+V_*\partial_y+V_*^2\nu_-^{-1}\partial_y^2+\nu_+\right]\phi\\
    -(2n_0)^{-1}\partial_t\delta n+(1-\eta_i)\frac{V_*}{\epsilon^{1/2}}\partial_y(\phi^2)\\
    +2\epsilon^{-1/2}\nu_-^{-1}V_*^2(\partial_x\ln n_0)^{-1}(\partial_y\phi\partial^2_{yx}\phi-\partial_x\phi\partial_y^2\phi)=0
\end{split}
\end{equation}
The nonlinearities including radial derivatives are smaller by $k_xk_yV_*\partial_x\ln (n_0)/\nu_-$, so we will ignore them. As pointed out in \cite{Diamond_Trapped_ion}, despite these terms being of lower order, neglecting them restricts the applicability of the model, due to removing a second dimension which the system could relax into. However, in the interest of simplicity of presentation, only the first two orders will be kept. 

Finally, as shown in Cohen et.al.\cite{Cohen_1978} Fourier-Laplace transforms along with a linear susceptibility relation determines the linear kinetic corrections
$$\delta \tilde{n}=-k^2\delta\chi\tilde{\phi}/4\pi e$$
Where $\delta\chi$ is obtained through the difference of fluid and kinetic dispersion relations. Ignoring finite banana-width effects, and remembering that the fast bounce motion of trapped ions approximately averages out magnetic shear effects, the only kinetic effect considered is ion Landau damping, whose contribution can be shown\cite{Cohen_1978} to be 
\begin{equation}\label{kineff}
    \delta n\approx -2n_0A'(1-3\eta_i/2)\omega_{Bi}^{-3}(V_*)^4\partial_y^4\partial_t^{-1}\phi
\end{equation}
Where $A'$ is a numerical factor\cite{Cohen_1976}. Combining \ref{2DKSE}, \ref{kineff}, ignoring the nonlinearities containing radial derivatives, and considering the maximally unstable regime of negligible ion-ion collisions ($\nu_+\sim 0$) yields a reduced, 1D model for the dissapative trapped ion mode:
\begin{equation}\label{dimrestKSE}
\begin{split}
    \left[\partial_t+V_*\partial_y+\frac{V_*^2}{\nu_-}\partial_y^2+A'(1-3\eta_i/2)\frac{V_*^4}{\omega_{Bi}^3}\partial_y^4\right]\phi\\
    +(1-\eta_i)\frac{V_*}{\epsilon^{1/2}}\partial_y(\phi^2)=0
\end{split}
\end{equation}
Shifting into the co-moving frame traveling at the trapped-electron diamagnetic drift velocity: $\eta=y-V_*t$, and using dimensionless variables:
$$\tau\equiv\frac{\omega_0^2t}{\nu_-},\hspace{3pt}\zeta\equiv\frac{\eta}{r},\hspace{3pt}\psi\equiv\frac{1}{2}\frac{\nu_-}{\omega_0}\frac{1-\eta_i}{\epsilon^{1/2}}\phi$$
$$\omega_0\equiv\frac{V_*}{r},\hspace{3pt}\nu\equiv A'\left(1-\frac{3\eta_i}{2}\right)\left(\frac{\omega_0}{\omega_{Bi}}\right)^2\frac{\nu_-}{\omega_{Bi}}$$
Yields the Kuramoto Sivashinsky equation
\begin{equation}\label{KSE}
    \begin{split}
        \partial_{\tau}\psi+\partial_\zeta^2\psi+\nu\partial_\zeta^4\psi+\frac{1}{2}\partial_\zeta(\psi^2)=0\\
        \psi(\zeta)=\psi(\zeta+2\pi)
    \end{split}
\end{equation}

The effective Reynolds number $\nu$ can be absorbed into the length and time scales:
$$x\equiv\nu^{-1/2}\zeta,\hspace{5pt}t\equiv\nu\tau,\hspace{5pt}u\equiv\nu^{-1/2}\psi$$
This scaling allows the maximum system size to be determined by Landau damping considerations. Here, we use $\nu\sim10^{-3}\rightarrow L_{max}\sim128$ as a rough estimate for DIII-D parameters. Using these scalings and doubly periodic boundary conditions, \eqref{dimrestKSE} becomes

\begin{equation}\label{MattKSE}
    \begin{split}
    \partial_tu+\partial^2_xu+\partial_x^4u+\frac{1}{2}\partial_x(u^2)=0\\
    x\in[0,L],\hspace{3pt} t\in[0,T]
    \end{split}
\end{equation}

Note that we have reused the coordinates $x$ and $t$ in an effort to match the notation commonly seen in the dynamical systems literature\cite{KSEState}. With doubly periodic boundary conditions, this reduced TIM model has been recast into a spatiotemporal field theory with the symmetry group of Galilean invariance and $G\equiv\text{SO}(2)\times \text{O}(2)$ (rotations in time, rotations and parity in space). The time and space domains have been have been left ambiguous as the goal is to find multiperiodic solutions of \eqref{MattKSE} with various spacetime volumes for use within the framework of multiperiodic orbit theory, whose tenants will be explained next.

\section{Periodic Orbit Theory}\label{third}
Periodic orbit theory is the mathematical framework which allows for quantitative predictions to be built off the back of periodic solutions to a given governing equation. Due to the data-driven nature of this current work, only the heuristic 
tenants of 
periodic orbit theory, and its extension to 
multi-periodic orbit theory, need be covered. For dissipative systems, small-scale 
fluctuations are damped out, forcing long-time state space dynamics to 
collapse to a finite dimensional inertial manifold 
$\mathcal{M}$\cite{ashtari_adjoint-based_2024, ding_estimating_2016}. 
Steady-state turbulence is confined to a "strange attractor" 
(also called a 
"chaotic attractor") $\mathcal{A}\in\mathcal{M}$. Broadly, a strange 
attractor is a compact attracting set with sensitivity to initial 
conditions\cite{ruelle_strange_1980}. In other words, the autocorrelation function 
for chaotic trajectories within $\mathcal{A}$ has finite support. 

There exists a dense set of unstable periodic solutions to the system's 
governing equations within the strange 
attractor\cite{On_Devaney's,Devaney_1989}. 
For autonomous (time-translation invariant) 
dynamical systems, these periodic orbits form a rigid road map to which 
chaotic solutions must adhere to\cite{ChaosBook}. The 
basic idea of periodic orbit theory is 
that each periodic orbit forms a saddle point where the stable manifold 
coaxes a chaotic trajectory towards resembling ("shadowing") the orbit, 
and the unstable manifold pushes it away to ensure the trajectory remains 
chaotic. As these periodic orbits are dense in the strange attractor, and 
each one is capable of forcing a chaotic trajectory to shadow, they can 
be used as support when taking averages, or as arbitrarily accurate 
initial guesses for general motion. Though, it is important to keep in mind that the more complicated the orbit, the more difficult it is for a trajectory to shadow (complicated orbits are more unstable), thus when taking averages the simplest orbits are the most important.

The natural extension of periodic orbit theory is to systems with 
multiple translational symmetries. In doing so, we switch 
from a "dynamical" perspective, to a "field-theoretic" perspective. Every 
translationally symmetric direction is treated equally, and the system's 
governing equation changes its role to the "law" which must be obeyed at 
every point in spacetime. Much like when only considering time, 
multi-periodic orbit theory builds a repertoire of solutions 
periodic in multiple directions; this allows for coverage in 
systems where the large spatial extent weakens spatial correlations between local structures that make up turbulent motion. Shadowing, and 
the ability to take averages remain in the extension to spacetime\cite{Cvitanovic2025chaotic}. 
Additionally, as will be discussed in the next section, the shift in 
perspective to field theory is computationally far preferable to 
traditional dynamical evolution.

\section{Numerical Framework for Extracting Coherent Structures}\label{fourth}


The exact coherent structures resulting from \eqref{MattKSE} are found using the \textit{Orbithunter} code developed in 2020 \cite{GudorfThesis} for the express purpose of determining spatiotemporal solutions of nonlinear PDEs. Only an outline of the numerical methods employed will be presented here as the full details are collected in the Ph.D. thesis that presented the code.

\textit{Orbithunter} constructs a discretized cost function $\phi(v)$, which is chosen to be one-half the squared error where the roots $\phi(v) = 0$ correspond to spatiotemporally periodic solutions. These roots are solved for in two stages: a least-squares optimization is solved to bring a randomly generated initial state closer to a solution and then a root-finding problem is solved using exact or inexact Newton methods. In the context of the Kuramoto Sivashinsky equation we can represent the collection of optimization variables as $v=[u,p]^T$. $v$ is a vector of the discretized potential fluctuation field, $u$, along with the parameters, $p$, which satisfy \eqref{MattKSE}. These parameters include the spatial extent, $L$; the temporal extent, $T$; and $S$, a "slant" parameter that accounts for \textit{relative periodic orbits}--orbits that are only periodic after a time $T$ when they are subjected to a spatial shift. Allowing each of these three parameters to vary simultaneously, is a technique that seems to be largely absent from existing literature. Most likely, this is due to the significant numerical challenge it presents. We can justify this formulation by defining a formal Lagrangian through an adjoint field $\lambda$\cite{ibragimov_lagrangian_2004}
\begin{equation}\label{lagrangian}
    \mathcal{L}=\frac{1}{2}\lambda\left[\partial_tu+\partial^2_xu+\partial_x^4u+\frac{1}{2}\partial_x(u^2)\right]
\end{equation}
From here, \eqref{MattKSE} can be represented by the function $f(v)$, which is identically zero when the RHS of the Kuramoto Sivishinsky equation is zero and all boundary conditions are satisfied with the given parameters ($L$, $T$, and $S$). Choosing $\lambda=f$ means that when \eqref{MattKSE} is satisfied, the adjoint equation is trivially satisfied. In doing so, the action
$$\mathcal{S}=\int_{\Omega}\mathcal{L}dtdx=\int_{\Omega}\frac{1}{2}|f|^2dtdx$$
has a global minimum at $\mathcal{S}=0$ only when $f=0$ identically throughout the tile. Requiring $f=0$ automatically defines the cost function which \textit{Orbithunter} minimizes:
\begin{equation}\label{costfunc}
    \phi(v)=\frac{1}{2}f^Tf=0
\end{equation}
Where, to reiterate, $f=f(v)$ is the functional representation of both the partial differential equation, and the parameters included in the solution.   
\subsection{Numerical Formulation} 
In order to frame \eqref{costfunc} as a numerical minimization problem we first assume doubly-periodic boundary conditions. \eqref{MattKSE} is transformed to a Fourier basis which results in a set of differential algebraic equations, which are then solved via pseudospectral methods.
\begin{equation}\label{pseudospecKSE}
\begin{split}
    f(v)=\left[\textbf{M}[\partial_t]+\textbf{M}[\partial^2_{x}]+\textbf{M}[\partial^4_{x}]\right]\tilde{u}+\\
    \frac{1}{2}\textbf{M}[\mathcal{F}_t\partial_x\mathcal{F}_x](\mathcal{F}^{-1}(\tilde{u})\cdot\mathcal{F}^{-1}(\tilde{u}))    
\end{split}
\end{equation}
Where $\textbf{M}$ denotes the matrix equivalent of an operator, $\tilde{u}$ is the Fourier modes of the potential fluctuations, and $\mathcal{F}$ is a spatiotemporal discrete Fourier transform\cite{GudorfThesis}. When searching for relative periodic orbits we use a different ansatz; $f_S$, defined as
\begin{equation}\label{co-moving}
    f_S=f(v)-\frac{S}{T}\textbf{M}[\partial_x]\tilde{u}
\end{equation}
This explicitly includes another optimization variable that accounts for spatial shifts.

Doubly periodic spatiotemporal patterns are found by numerically optimizing either \eqref{pseudospecKSE} or \eqref{co-moving}. As this problem is framed as a variational descent with no exponential Lyapunov divergence, we generate initial conditions for the Fourier modes using modulated random noise. The first step is to create a collection of periodic orbits resulting from aforementioned process. Identification of "fundamental orbits" is then accomplished by "clipping" out patterns that frequently occur in this set. These clippings are then used as initial conditions in a second round of numerical optimization. After the set of fundamental orbits have been obtained (the requirements for being in such a set will be laid out shortly), members of the set can be "glued" together and used as initial guesses for large orbits with particular attributes. In this way, the turbulent dynamics of the TIM has been reduced to an ordered combination of fundamental behaviors embodied by recognizable, coherent structures. The following subsections explore each step of this process in detail.          

\subsubsection{Numerical Optimization Method}
Solving \eqref{MattKSE} reduces to applying minimization algorithms to \eqref{costfunc}. While there are an infinite number of ways of doing this, we have found the following to be a consistent setup. 

Firstly, we execute the adjoint descent method\cite{Farazmand2016}, which evolves the cost function in fictitious time enforcing $\phi(v+(\partial_{\tau}v)\delta \tau)\leq\phi(v)$. Differentiating the cost function \eqref{costfunc} with respect to fictitious time, the minimization is guaranteed by requiring 
\begin{equation}\label{descent}
    \partial_{\tau}\phi=[\mathbf{J}\partial_{\tau}v]^Tf<0
\end{equation}
Setting 
\begin{equation}\label{adjointevol}
    \partial_{\tau}v=-\mathbf{J}^Tf
\end{equation}
Automatically satisfies \eqref{descent} and serves as the update equation for $v$. After descending past a certain tolerance, further refinement is achieved by applying an inexact Newton's method, which solves the Newton's equation least-squares problem $\mathbf{J}\delta v\approx-f$ via pseudoinverse\cite{virtanen_scipy_2020}. As solving \eqref{MattKSE} demands optimization of both the potential field and the system parameters, the Jacobian includes partial derivatives with respect to said parameters 
\begin{equation}\label{jacobians}
    \begin{split}
        \mathbf{J}\equiv\frac{\partial f}{\partial v}=\left[\frac{\partial f}{\partial \tilde{u}},\frac{\partial f}{\partial L},\frac{\partial f}{\partial T}\right],\\
        \mathbf{J}_S\equiv\frac{\partial f_S}{\partial v}=\left[\frac{\partial f_S}{\partial \tilde{u}},\frac{\partial f_S}{\partial L},\frac{\partial f_S}{\partial T},\frac{\partial f_S}{\partial S}\right]
    \end{split}
\end{equation}
The adjoint Jacobian used for the adjoint descent method \eqref{adjointevol} is derived through linearizing the adjoint equation with respect to $\lambda=f$ from \eqref{lagrangian}:
\begin{equation}
    \begin{split}
        \frac{d\mathcal{L}}{\partial u}=\left(\frac{\partial}{\partial u}-\frac{\partial}{\partial x}\frac{\partial}{\partial u_x}-\frac{\partial}{\partial t}\frac{\partial}{\partial u_t}+\frac{\partial^2}{\partial x^2}\frac{\partial}{\partial u_{xx}}\right.\\
        +\left.\frac{\partial^4}{\partial x^4}\frac{\partial}{\partial u_{xxxx}}\right)\mathcal{L}
        =-\partial_tf+\partial_x^2f+\partial^4f-u\partial_xf\\
    \end{split}
\end{equation}
The terms associated with solving for the problem parameters ($L$, $T$, $S$) remain the same as in \eqref{jacobians}. The members of the Jacobian are given explicitly in Appendix A.

\begin{figure}[ht]
\includegraphics[scale=0.35]{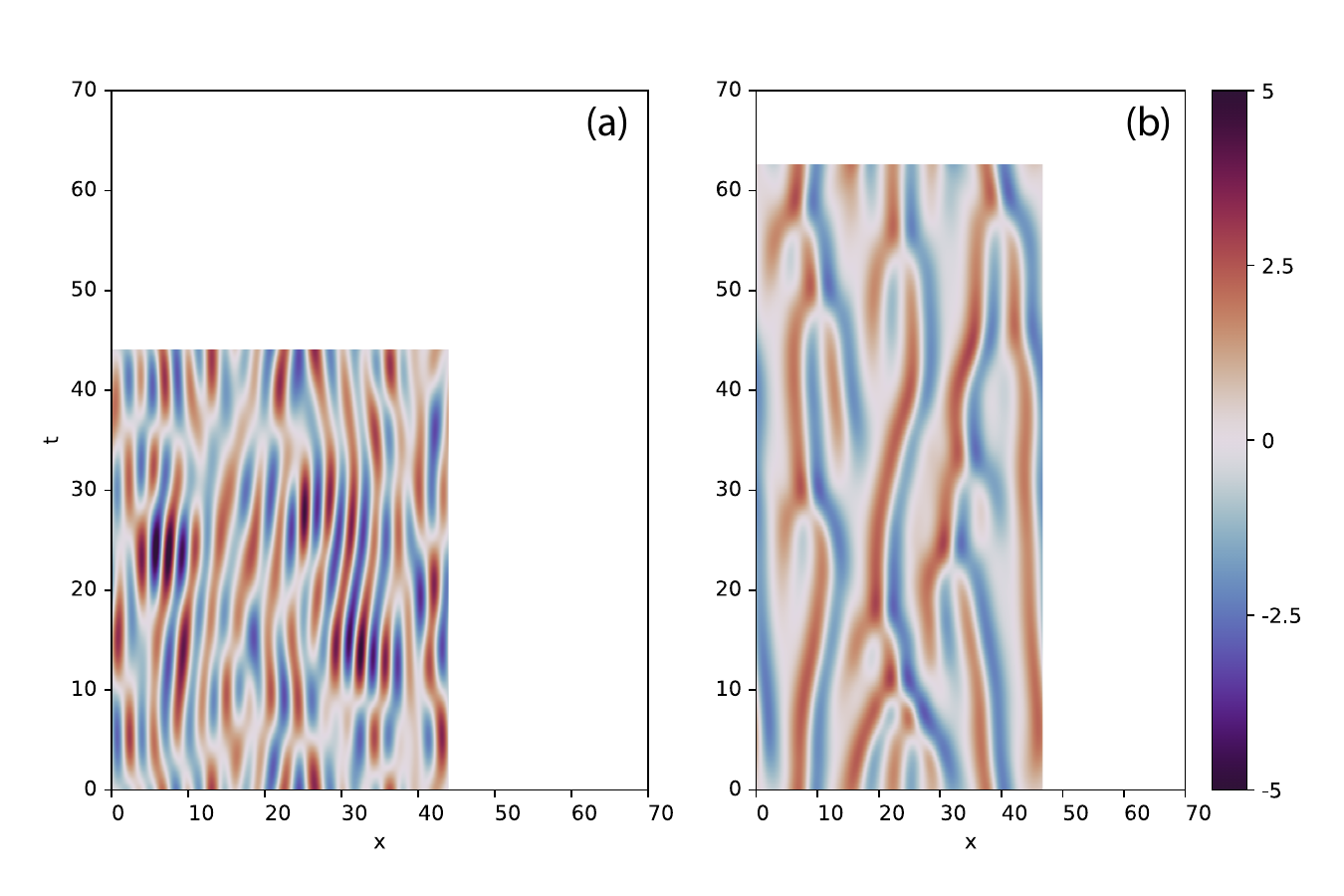}
\caption{\label{Fmodes} An example of converging a solution (b) of the Kuramoto Sivashinsky equation using random Fourier noise (a) as an initial guess. Note that due to optimization of system parameters, the spacetime volume of (a) is different than the spacetime volume of (b).}
\end{figure}

\subsubsection{Clipping: Identifying Fundamental Periodic Orbits}
\begin{figure}[ht]
\includegraphics[scale=0.45]{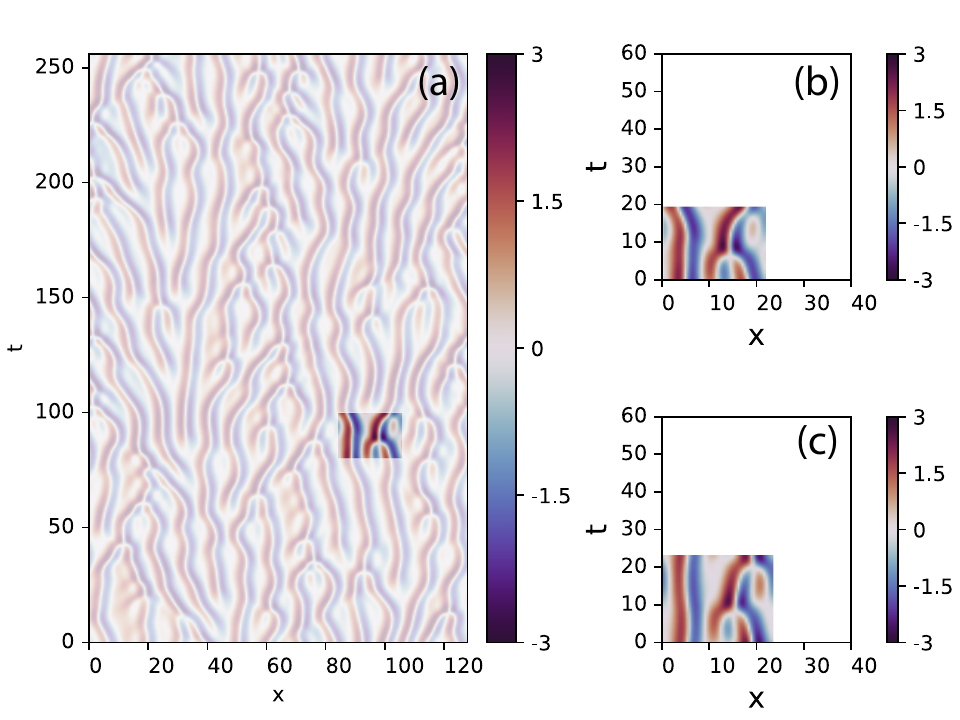}
\caption{\label{clip} An example of clipping. (a) shows the large spacetime orbit, where the saturated square is the clipping that has been selected. (b) is the initial guess fed into the optimization algorithm, and (c) is the result of numerical optimization. Note that (b) and (c) are visually very similar, but still distinct, this implies (B) was closely shadowing (c).}
\end{figure}

After running guess orbits through numerical optimization, a collection of solutions to \eqref{MattKSE} is created. From here, it is possible using \textit{Orbithunter's} "clipping" functionality to isolate smaller solutions by selecting windows of spacetime in converged solutions and using them as initial guesses (FIG. \ref{clip}). 

\begin{figure}[ht]
\includegraphics[scale=0.55]{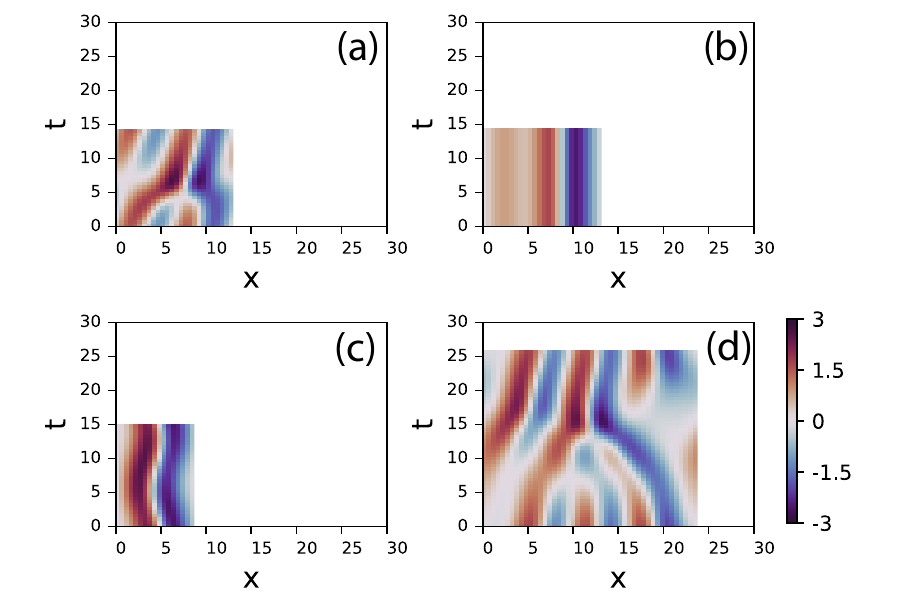}
\caption{\label{FPOs} The fundamental periodic orbits of the Kuramoto Sivashinsky equation. (a) encodes wave-wave interaction and is a relative periodic orbit. (b) is the equilibrium, which has been given a well-defined spacetime volume. (c) encodes simple time dependence. And (d) is, the much less common three-wave interaction.}
\end{figure}

The goal in doing this is to find the set of smallest possible solutions which together display all fundamental behaviors of the system. These minimal solutions will hereafter be referred to as "fundamental periodic orbits". These are the smallest possible spacetime patterns with no internal or repeating structure that each display a certain fundamental behavior. Because these patterns are fundamental, they are highly recurrent. For the case of the TIM model in this paper, the "complete set of fundamental behaviors" was found to be the equilibrium given by a single peak-trough pair, a wave packet whose location oscillates in time, the most simple example of wave-wave interaction: two waves colliding and becoming one, and a large triad (three-wave) interaction. This set of fundamental periodic orbits is visualized in FIG. \ref{FPOs}

\subsubsection{Gluing: Constructing Larger Spatiotemporal Patterns Using Fundamental Periodic Orbits}
\begin{figure}[h]
\includegraphics[scale=0.35]{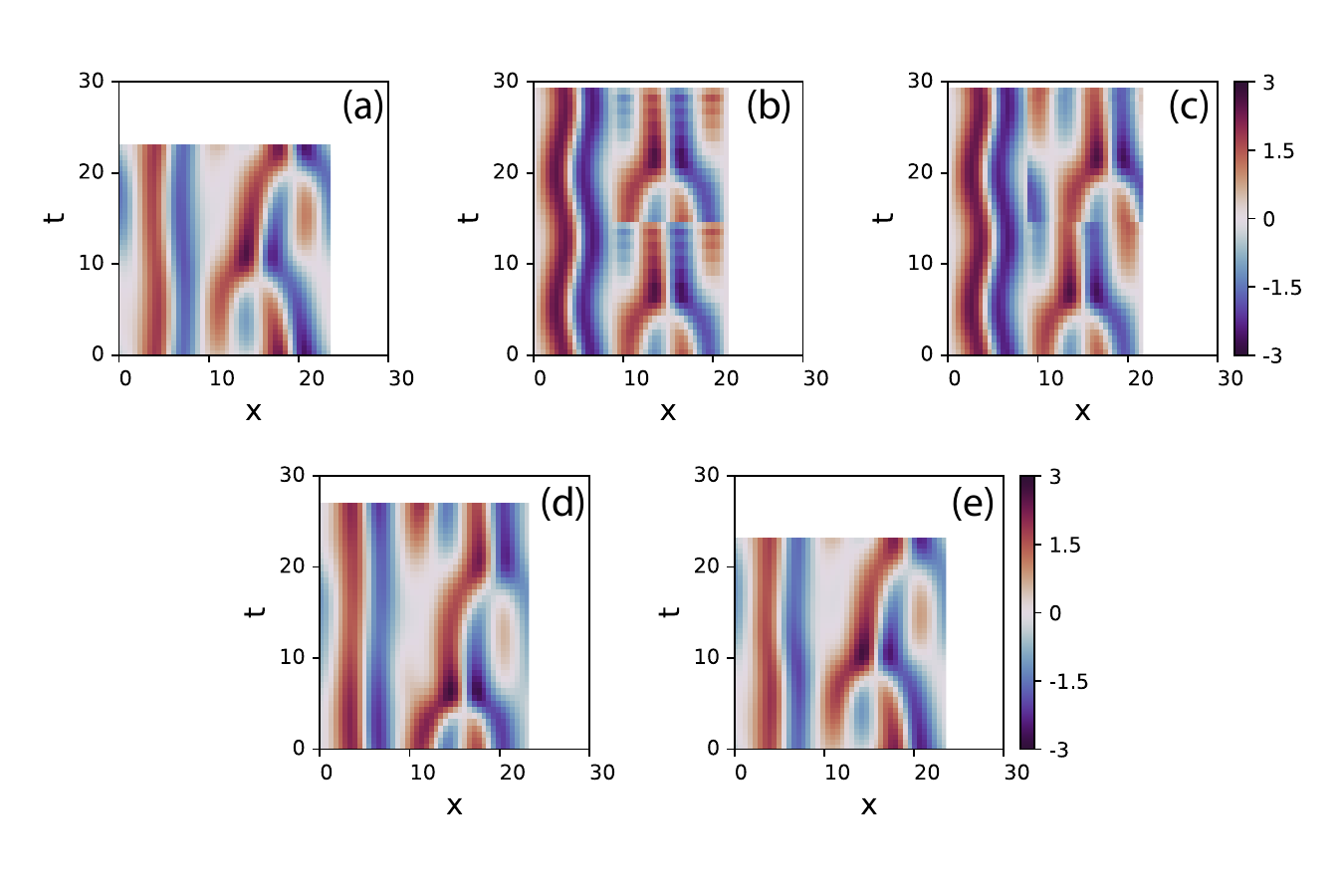}
\caption{\label{glue} A minimal gluing example. (a) is the reference orbit converged from clipping, (b) is the naive guess, (c) is the symmetry optimized guess, and (d) is the solution converged from (c). (e) is the result of applying numerical continuation to (e).}
\end{figure}

The inverse process of clipping, "gluing", takes small, simple solutions and combines them to create iteratively larger solutions. In practice, this equates to taking a collection of similarly sized converged spacetime patterns and applying postprocessing to each one--as detailed in \cite{GudorfThesis}--so as to make concatenation well defined. This larger concatenated array serves as a physics-informed initial guess for a new solution which is not only of a (approximately) specified spacetime volume, but also displays specified physical processes at its base (eg. some number of wave-wave interactions, or periods of local equilibrium). This is best seen through example. Taking the orbit converged from clipping in FIG. \ref{clip}, there appears to be two wave-wave interactions on the right, and a time-dependent peak-trough pair on the left. Using the labeling from FIG. \ref{FPOs} the heuristic behavior of the clipped orbit can be described as the following matrix:
\begin{equation}\label{GlueMAT}
    \begin{bmatrix}
        C & A \\
        C & A
    \end{bmatrix}
\end{equation}
Before using \eqref{GlueMAT} as an initial guess, spatial and temporal translation invariance can help reduce discontinuities. Since \eqref{MattKSE} is translation-invariant, any cyclic permutation of an orbit is equally valid (arbitrary continuous rotations induce interpolation error). This means the building blocks can (and should) be cyclically shifted to minimize artificial discontinuities which do not befit the smooth solutions associated with the highly dissipative equation. After converging the symmetry optimized guess, the spacetime volume can be set more precisely through numerical continuation (constraining one dimension and repeatedly reconverging the solution to explore its continuous families which are equivalent in the eyes of multiperiodic orbit theory)\cite{GudorfThesis}. This process is presented pictorially in FIG. \ref{glue}.

Though not the focus of this current work, gluing and clipping are excellent examples of how the concept of shadowing can be used at every level of analysis of complex nonlinear systems. Using fundamental solutions to pinpoint a larger, more complex, solution is better informed than starting with initial noise or a time integrated recurrences.

\subsection{Shadowing and Orbit Statistics}
The power of using periodic orbits to investigate chaotic motion lies in "shadowing"; the idea that an arbitrary pattern will inevitably come close to the periodic orbits dense in state space. This is central to both clipping and gluing, as each rely on the assumption that large spacetime solutions of \eqref{MattKSE} consist of many close passes of simpler patterns. Gluing could not converge solutions if ensembles of small periodic orbits were not good guesses for large orbits, and clipping could not converge small orbits if large orbits did not consist of many close approximations of simple solutions. As shadowing naturally manifests as small solutions recurring in the broader spacetime it is possible to conduct global averages of observables by finding the observable average on individual orbits and then carrying out a weighted sum (see Appendix B for details). In \textit{Orbithunter} the weights in this sum are computed through successive numerical experiments, where each fundamental orbit is "scored" through how often they appear in a large sample set of long time-integrated trajectories generated via random initial conditions. 

The scoring is conducted through a process similar to a convolution, except the operation applied within the "convolution" window is an $L2$ norm. Each pixel of a time integrated solution is scored for each fundamental orbit using the following formula 
\begin{equation}\label{L2convo}
    S(i,j)\equiv\frac{1}{n*m}\sqrt{{\sum^{m-1}_{p=0}\sum^{n-1}_{q=0}}\left(\tilde{A}(i+p,j+q)-K(p,q)\right)^2}
\end{equation}
Here, $\tilde{A}=A-\overline{A}$ is the underlying snippet of a large solution normalized such that the local (within the "convolution" window) mean flow is zero. Normalizing like this accounts for the global (spacetime) average of potential fluctuations being zero. $K$ is one of the fundamental periodic orbits, with $m$ spatial nodes, and $n$ temporal nodes. 

\begin{figure}[ht]
\centering
\begin{subfigure}{0.45\textwidth}[\rom{1}]
    \centering
    \includegraphics[scale=0.35]{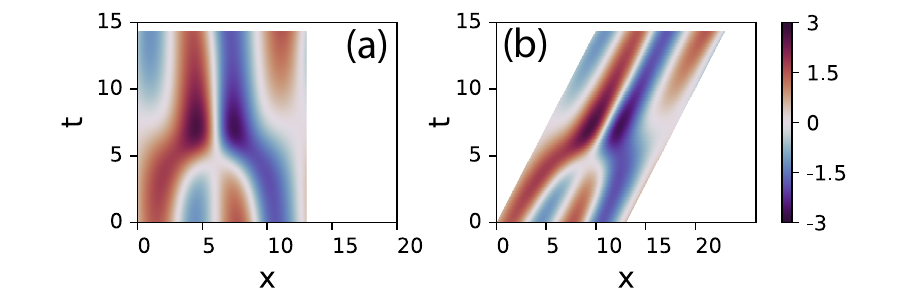}
    \caption{\label{parapip} An example of projecting the wave-wave interaction (a) into a local co-moving frame (b). This allows for more direct comparison with large time-integrated solutions that have local pockets of non-zero average velocity.}
\end{subfigure}
\vfill
\begin{subfigure}{0.45\textwidth}[\rom{2}]
    \centering
    \includegraphics[scale=0.35]{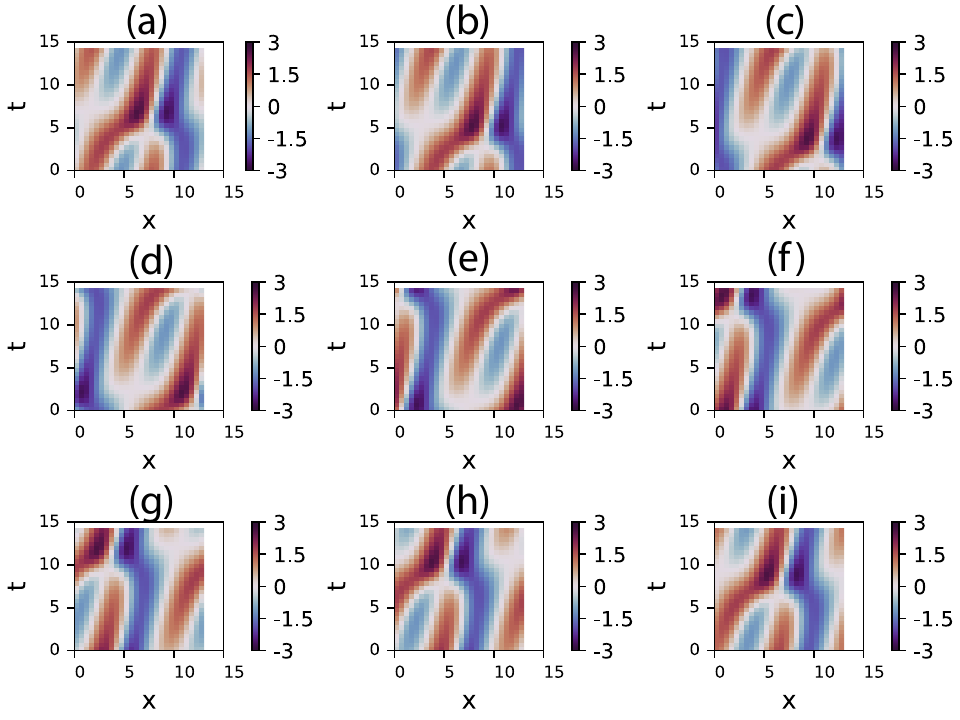}
    \caption{\label{permut} A sample of the cyclic permutation $(C_n \times C_m = Time \times Space)$ symmetry group representations of the wave-wave interaction projected into its global co-moving frame. The global co-moving frame is the representation of the wave-wave interaction where it is a periodic orbit with non-zero average velocity rather than a relative periodic orbit.}
\end{subfigure}
\end{figure}

When calculating scores, the symmetry of \eqref{MattKSE} must be taken into account. The reasons for this is that the L2 norm is not invariant under the same group of transformations. Topological or symmetry invariant metrics can easily be argued for, but the L2 norm suffices for a first order calculation. By definition, a solution to an equation is still a solution after being acted upon by a member of the equation's symmetry group. Therefore, each member of a fundamental periodic orbit's "group orbit"--the set of state space points into which a fundamental orbit is mapped to under the action of the Kuramoto-Sivashinsky's symmetry group\cite{ChaosBook}--is an equally valid representation. For the discretized version of \eqref{MattKSE}, the symmetry group \( G \equiv \text{SO}(2) \times \text{O}(2) \) transforms into a set of discrete symmetries. These include cyclic permutations in both the space and time directions (see FIG. \ref{permut}), as well as spatial parity, given by \( -u(-x,t) \). The effect of parity is seen through mirroring the shape in FIG. \ref{FPOs}C and multiplying the result by $-1$ restore full multiperiodicity. The Galilean invariance is addressed by projecting into either local or global co-moving frames. Local frames are defined by pockets of a large solution, while global frames correspond to a nonzero average velocity for the entire solution (see FIG. \ref{parapip}). We took advantage of this symmetry to assign a definite spacetime volume to our time equilibrium (see FIG. \ref{FPOs}A). 

\begin{figure*}[ht]
    \centering
    \setlength{\abovecaptionskip}{0pt}  
    
    \begin{subfigure}{0.48\textwidth}
        \centering
        \includegraphics[height=5cm]{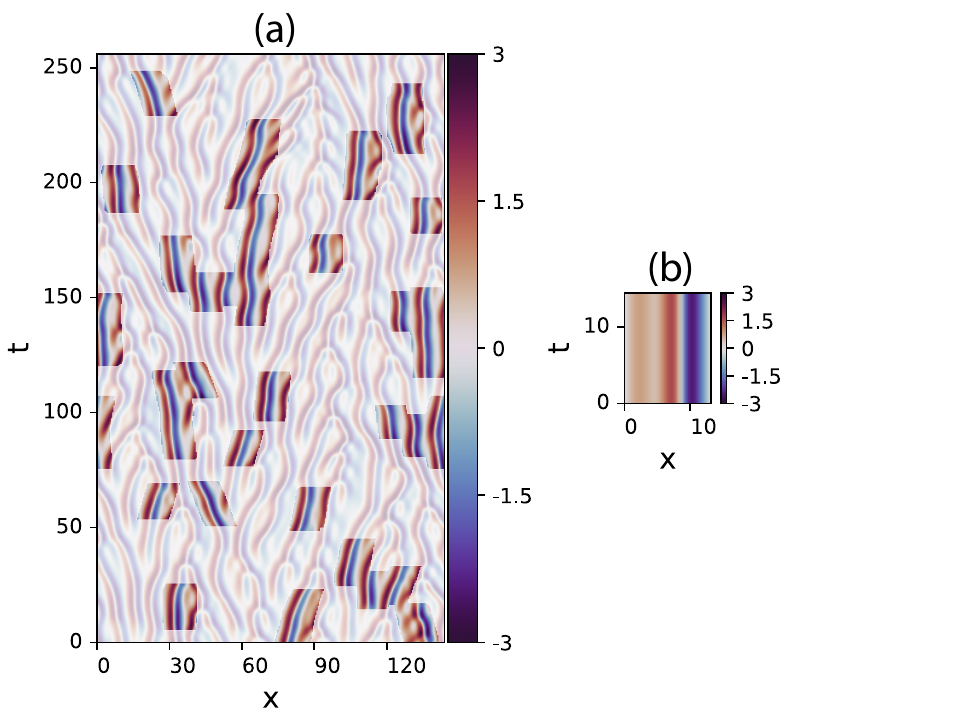}
        \caption{\label{cov0} Shadowing of the relative  equilibrium solution (b) depicted in co-moving frame}
    \end{subfigure}
    \hfill
    \begin{subfigure}{0.48\textwidth}
        \centering
        \includegraphics[height=5cm]{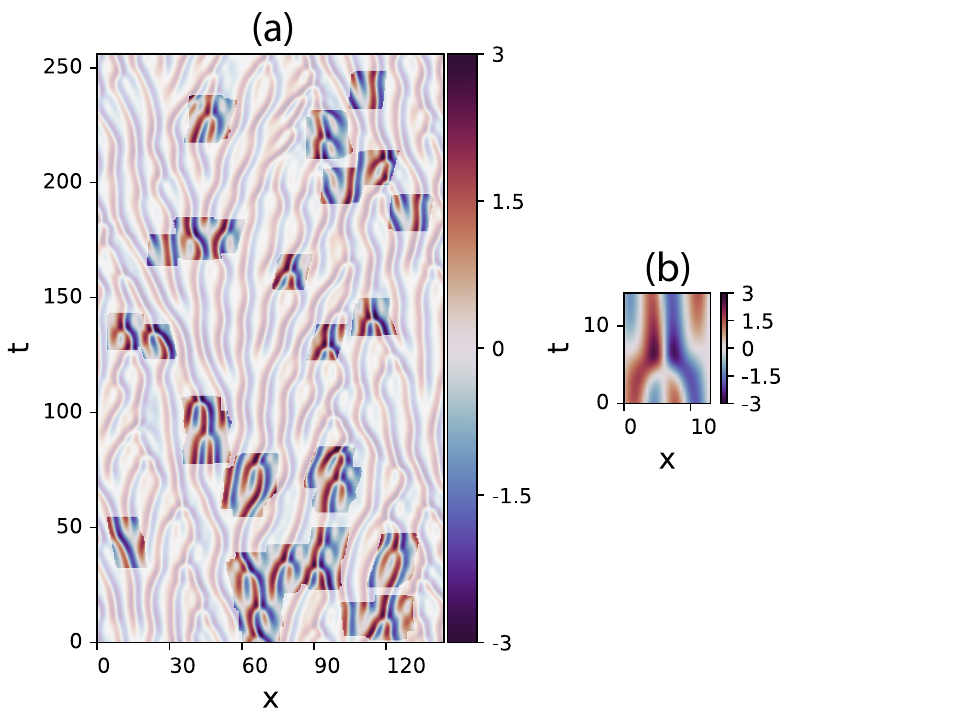}
        \caption{\label{cov1} Shadowing of the wave-wave interaction in the "lab frame" (b)\vspace{0.38cm}}
    \end{subfigure}

    \vspace{0.3cm}
    
    \begin{subfigure}{0.48\textwidth}
        \centering
        \includegraphics[height=5cm]{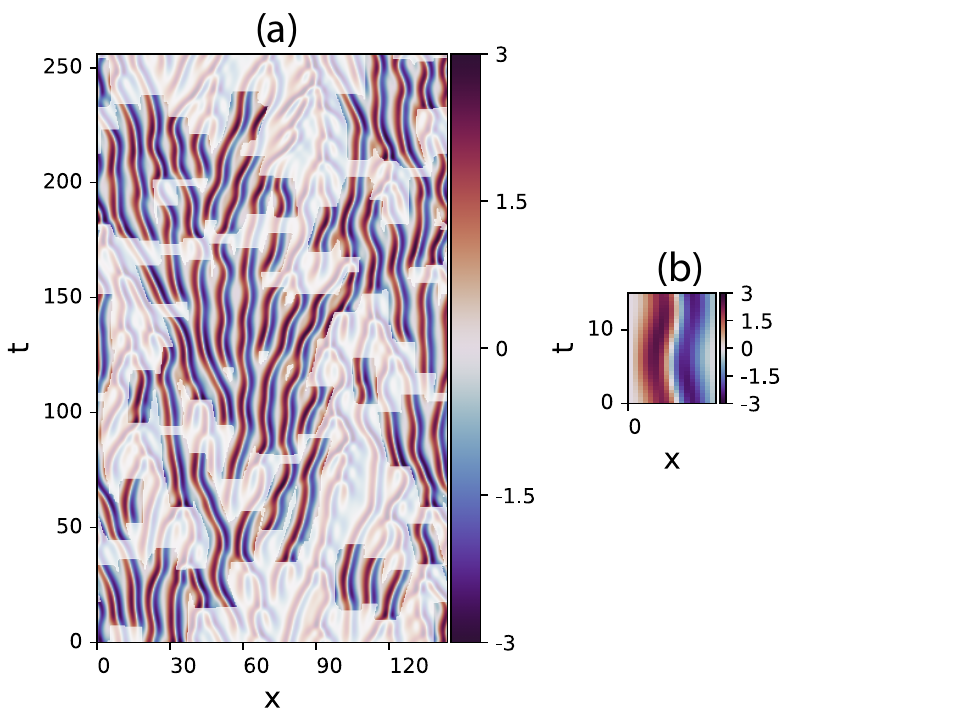}
        \caption{\label{cov2} Shadowing of the time-dependent wave (b)\vspace{0.88cm}}
    \end{subfigure}
    \hfill
    \begin{subfigure}{0.48\textwidth}
        \centering
        \includegraphics[height=5cm]{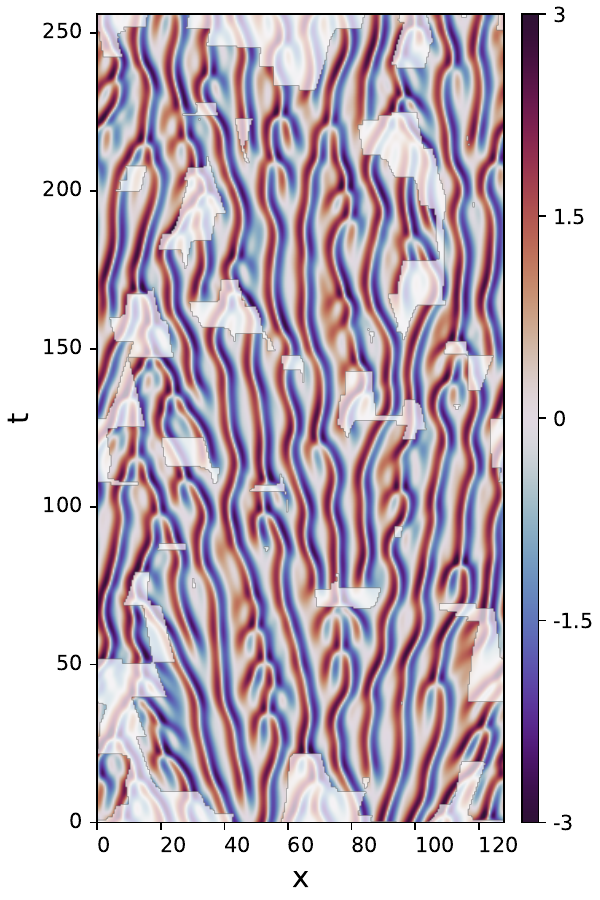}
        \caption{\label{cov9} Overall shadowing accounting for all fundamental periodic orbits and their group orbits. Washed out space is where shadowing was not detected.}
    \end{subfigure}

    \vspace{0.2cm}
    
    \caption{\label{covering} Shadowing coverage within a flow generated by integrating \eqref{MattKSE} using a random initial condition (shown as (a) in each subfigure) for each fundamental periodic orbit, aside from the triad whose spacetime volume is so large that shadowing is very infrequent. \ref{cov9} accounts for parity and global co-moving frames.}
\end{figure*}

The score of a given fundamental orbit is calculated by plugging its group orbit and all projections into local co-moving frames into \eqref{L2convo} for each pixel of a large sample set of time integrated solutions initialized with random initial conditions. When all scores have been calculated, they are compared pixel-by-pixel and the fundamental orbit which has the smallest score "wins" the pixel as long as the score passes beneath a certain threshold. Here, we use $S(i,j)<0.018$ as the threshold, well below the "visual" limit determined in experimental investigations of shadowing \cite{suri_capturing_2020}, and empirically found to minimize false positives. The shadowing detections for each fundamental periodic orbit, accounting for parity and global co-moving frames, are depicted in FIG. \ref{covering}.

The weights used for each fundamental orbit in the weighted averages of observables are then estimated by 
$$\text{weight}\sim\# \text{ of pixels won}/\#\text{ of pixels total}$$
Where $\#\text{ of pixels total}$ is the total number of pixels in a set of samples of forward in time integrated solutions to \eqref{MattKSE}. The number of samples is chosen such that the average observables explored in the next section converge to a steady value.

\section{Physical Observables from Fundamental Structures}\label{fifth}
 With weights in hand from shadowing calculations, it is possible to take advantage of the periodic points dense in the Kuramoto Sivashinsky strange attractor. The fundamental orbits gathered through clipping represent the least complicated (and therefore the least unstable and most important) doubly periodic solutions; they will serve as support for the averages taken in this section. The averaging computation manifests itself as a "typical" Birkhoff sum (for details see Appendix B):

 \begin{equation}\label{typical_Birk}
     \br{a}\approx\frac{w_1*A_1+w_2*A_2+w_3*A_3+w_4*A_4}{w_1*V_1+w_2*V_2+w_3*V_3+w_4*V_4}
 \end{equation}

Where the numerator of \eqref{typical_Birk} is a weighted sum--using the weights, $w_i$, found in the previous section--of some observable $A_i$ evaluated, and summed over the entire $i^{th}$ FPO, and the denominator is the "typical" spacetime volume of an FPO. 

 Although the expectation value of any physical observable could conceptually be taken, here only the potential fluctuation intensity, anomalous radial diffusion coefficient, wave energy dissipation, and radial turbulent energy flux will be found. Note that in all cases, the same set of weights are used. 

\begin{figure}[ht!]
    \centering
    \includegraphics[width=0.85\linewidth]{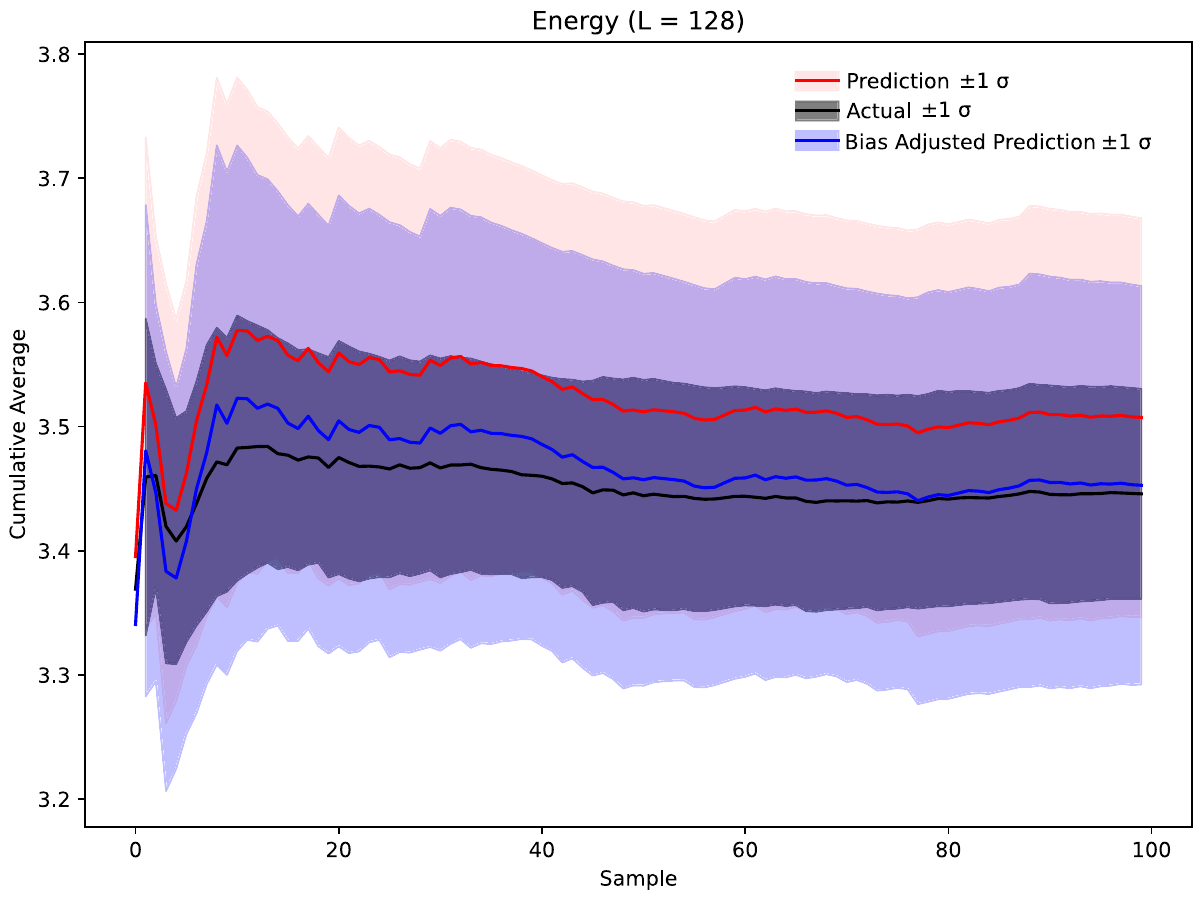}
    \caption{The average energy $u^2$ for $L=128$, found both through \eqref{typical_Birk} and averaging over many samples of solutions to \eqref{MattKSE}. The prediction line is calculated using the weights determined with the number of samples indicated on the x-axis. \eqref{typical_Birk} has an error of $1.8\%$. With bias correction, the error is $0.2\%$. In all cases, the shaded region is the $\pm 1\sigma$ range for the given set of samples.}
    \label{fig:avg_energy_128}
\end{figure}

 The average fluctuation intensity (or "energy"), $\br{u^2}$, predicted from \eqref{typical_Birk}, compared to the actual average from the many samples of flow generated from random initial seeds used to generate the weights, and a "bias-corrected" prediction explained below is presented in \ref{fig:avg_energy_128}.

\begin{figure}[ht!]
    \centering
    \includegraphics[width=0.85\linewidth]{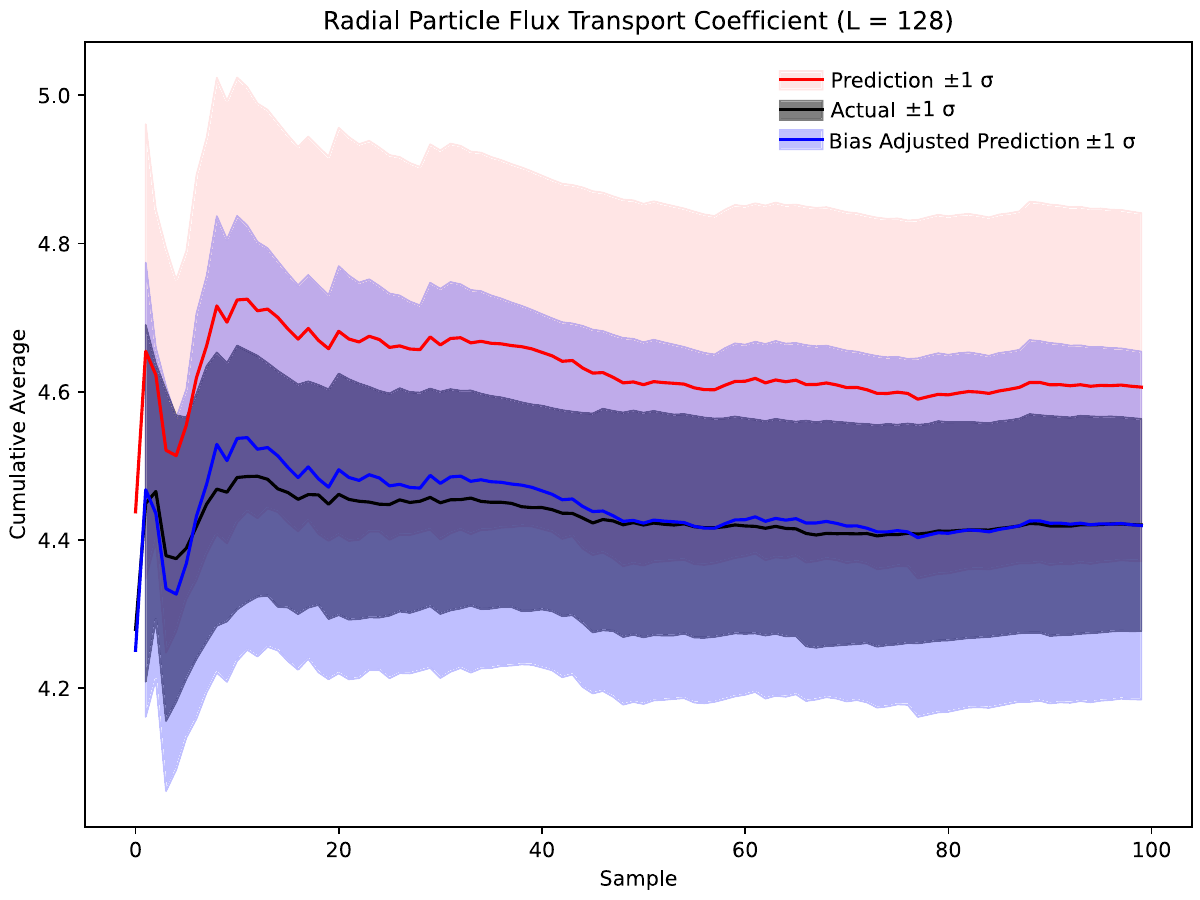}
    \caption{The average anomalous diffusion coefficient for $L=128$, found both through \eqref{typical_Birk} and averaging over many samples of solutions to \eqref{MattKSE}. The prediction line is calculated using the weights determined with the number of samples indicated on the x-axis. \eqref{typical_Birk} has an error of $4.2\%$. With bias correction, the error is $0.02\%$. In all cases, the shaded region is the $\pm 1\sigma$ range for the given set of samples.}
    \label{fig:avg_power_128}
\end{figure}

The functional form of the anomalous particle transport coefficient (going as $(\partial_y\phi)^2$) is found either by truncating electron density to the same order as \eqref{dimrestKSE}\cite{Cohen_1976}, or through application of quasilinear theory. Due to there being so much scaling involved when going from \eqref{dimrestKSE} to \eqref{MattKSE}, it is convenient to define a "scaled" transport coefficient which only depends on the explicit quantities being treated:
 \begin{equation}\label{anomD}
     D_s=\left\langle\left(\frac{\partial u}{\partial x}\right)^2\right\rangle,\quad D_s\propto D_{anomalous}
 \end{equation}
 Note that here, $\propto$ does not indicate strict proportionality, as $D_{anomalous}$ includes factors of $T$ that are not constant. However, since these factors are input parameters rather than quantities determined self-consistently within the model, they will be neglected in the subsequent calculations. The result of applying \eqref{typical_Birk}, and conducting the same comparison as in FIG. \ref{fig:avg_energy_128} is presented in FIG. \ref{fig:avg_power_128}.

\begin{figure}
    \centering
    \includegraphics[width=0.85\linewidth]{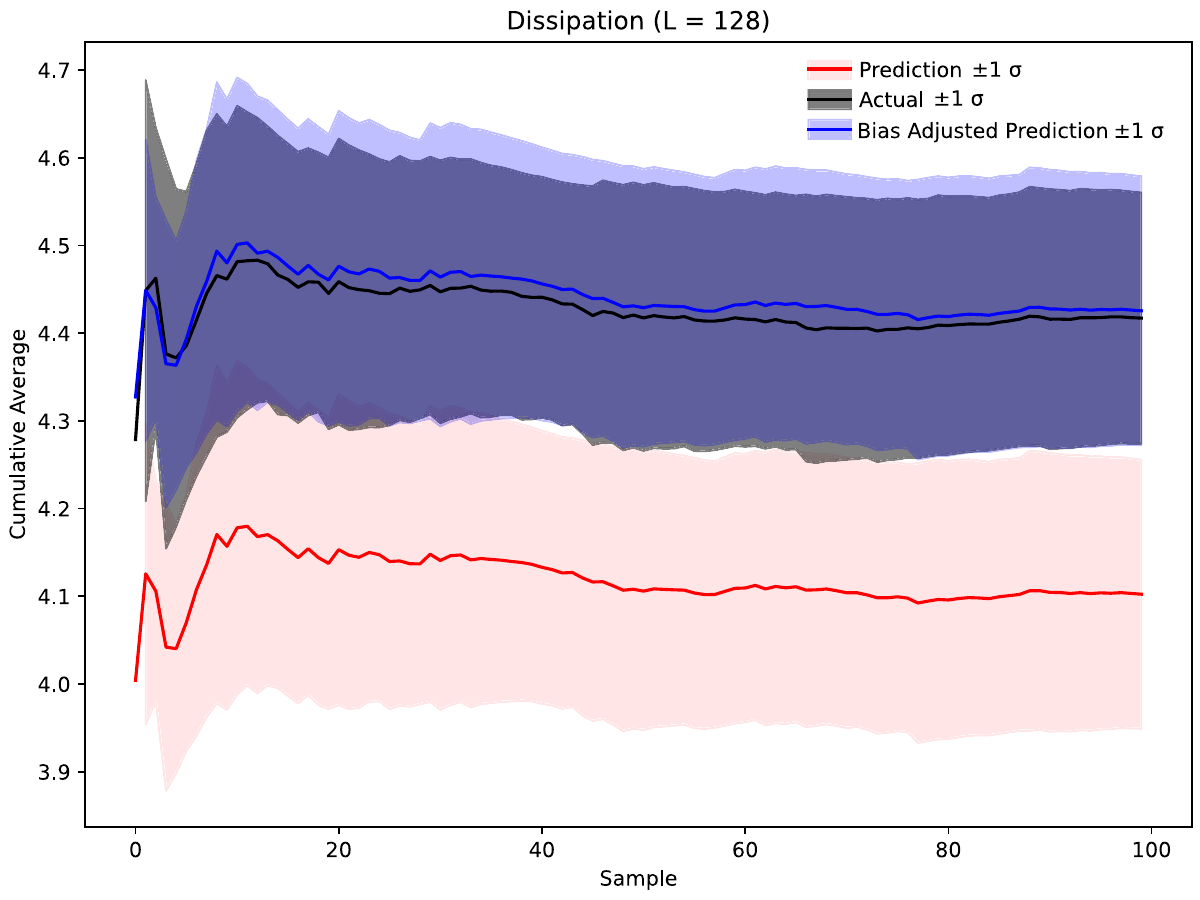}
    \caption{The average dissipation for $L=128$, found both through \eqref{typical_Birk} and averaging over many samples of solutions to \eqref{MattKSE}. The prediction line is calculated using the weights determined with the number of samples indicated on the x-axis. \eqref{typical_Birk} has an error of $7.1\%$. With bias correction, the error is $0.19\%$. In all cases, the shaded region is the $\pm 1\sigma$ range for the given set of samples.}
    \label{fig:avg_diss_128}
\end{figure}

In the Kuramoto Sivashinsky model for the dissipative TIM, energy cascades from long to short wavelengths where it is dissipated by Landau damping which is accounted for via the fourth order derivative in \eqref{MattKSE}\cite{laquey_nonlinear_1975}. As the spatial boundary conditions are periodic, averages of total space derivatives evaluate to zero: $\br{\partial_xf(u)}=0$. This allows the average dissipation to be written as the average squared second derivative:
\begin{equation}
    D=\br{\left( \frac{\partial^2 u}{\partial x^2}\right)^2}
\end{equation}
The same comparison as for the other observables is presented in FIG. \ref{fig:avg_diss_128}.

\begin{figure}
    \centering
    \includegraphics[width=0.85\linewidth]{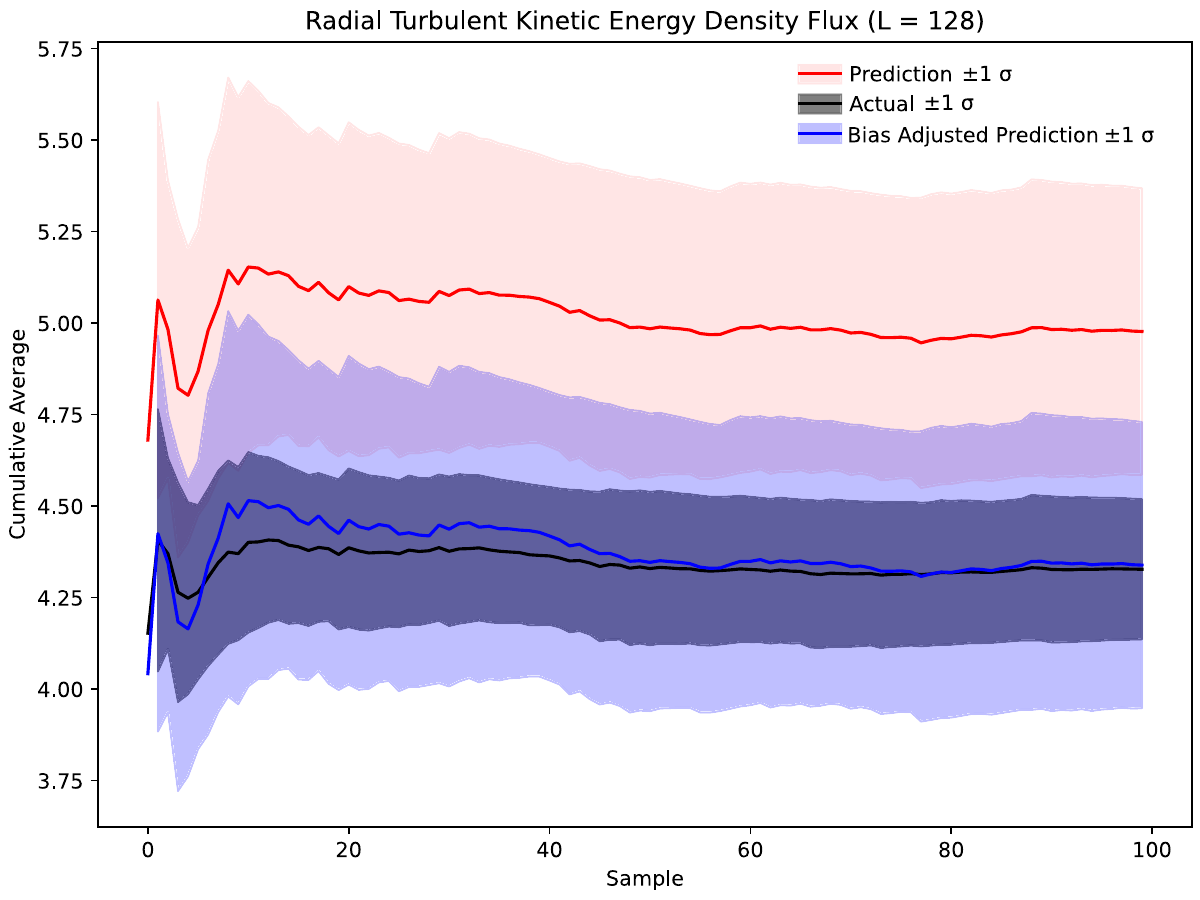}
    \caption{The average turbulent kinetic energy density flux for $L=128$, found both through \eqref{typical_Birk} and averaging over many samples of solutions to \eqref{MattKSE}. The prediction line is calculated using the weights determined with the number of samples indicated on the x-axis. \eqref{typical_Birk} has an error of $15\%$. With bias correction, the error is $0.26\%$. In all cases, the shaded region is the $\pm 1\sigma$ range for the given set of samples.}
    \label{fig:avg_turbkin_128}
\end{figure}

Finally, an example of an observable specifically relevant to modern plasma turbulence investigations\cite{Wu_etal_2023}: turbulent kinetic energy density flux. The dominant radial transport mechanism is the radial component of the $E\times B$ velocity generated by potential fluctuations of the TIM. This is proportional to $\partial_xu$ and so the scaled turbulence spreading to the same order as \eqref{dimrestKSE} is given by
\begin{equation}\label{kinfluxexpec}
    \left\langle v^2_{E\times B}\mathbf{v}_{E\times B}\cdot\hat{r}\right\rangle\propto-\left\langle\left(\frac{\partial u}{\partial x}\right)^3\right\rangle
\end{equation}
Where the overall minus sign is a result of accounting for the direction of flux. The same comparison as for the other observables is presented in FIG. \ref{fig:avg_turbkin_128}.

\begin{figure}
    \centering
    \includegraphics[width=0.95\linewidth]{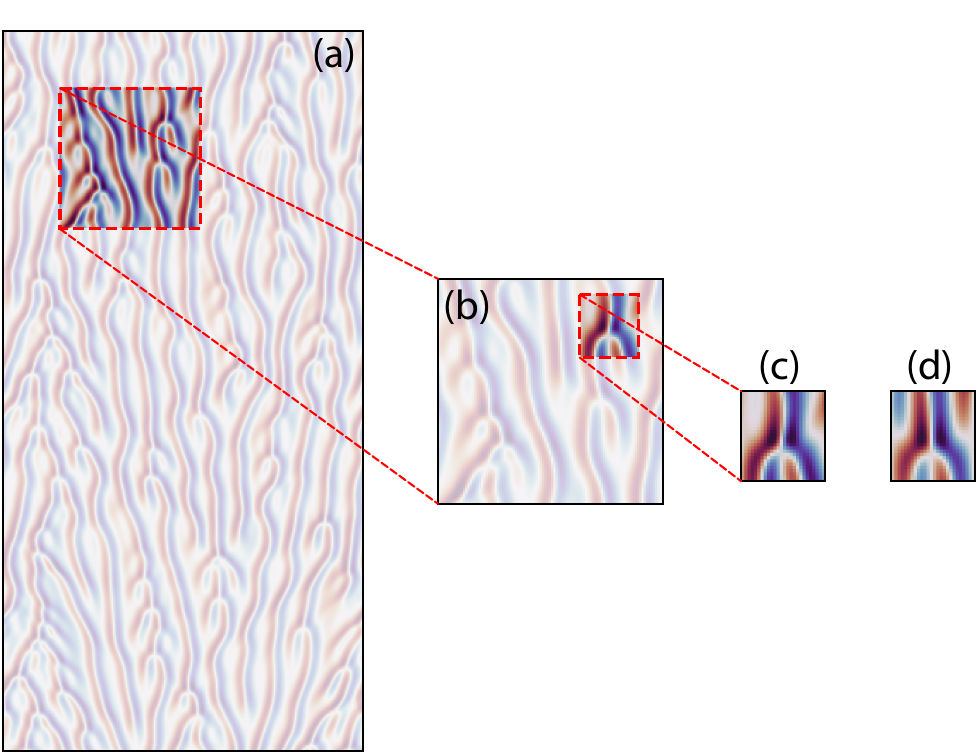}
    \caption{An example of a false negative during the shadowing calculation. Here, $S\sim0.0218$, and so (c) would not have been included as a shadowing event of (d) (one of the FPOs) when calculating weights.}
    \label{fig:failed_shadow}
\end{figure}

An unfortunate artifact of approaching this problem from a purely data analytics perspective, is that it is often impossible to construct perfect detection cutoffs. This is a result of the strong nonlinear coupling between neighboring field values, which distorts both the shape and magnitude of local patches in solutions to \eqref{MattKSE}. Our scoring parameter is a perfect example of this. As can be seen in FIG. \ref{fig:failed_shadow}, an obvious shadowing event can be passed over while avoiding the greater evil of large numbers of false positives. 

This changes the weights, and then affects the predicted observables. This becomes especially destructive for higher-order observables such as dissipation, and turbulent kinetic energy density flux that rely on accurate shadowing of the fine details of a flow. However, as this is a failing of computational cutoffs, instead of the theory itself, it is possible to treat the error simply as a constant bias. 

\begin{figure*}[ht]
    \centering
    \setlength{\abovecaptionskip}{0pt}  
    
    \begin{subfigure}{0.48\textwidth}
        \centering
        \includegraphics[height=6cm]{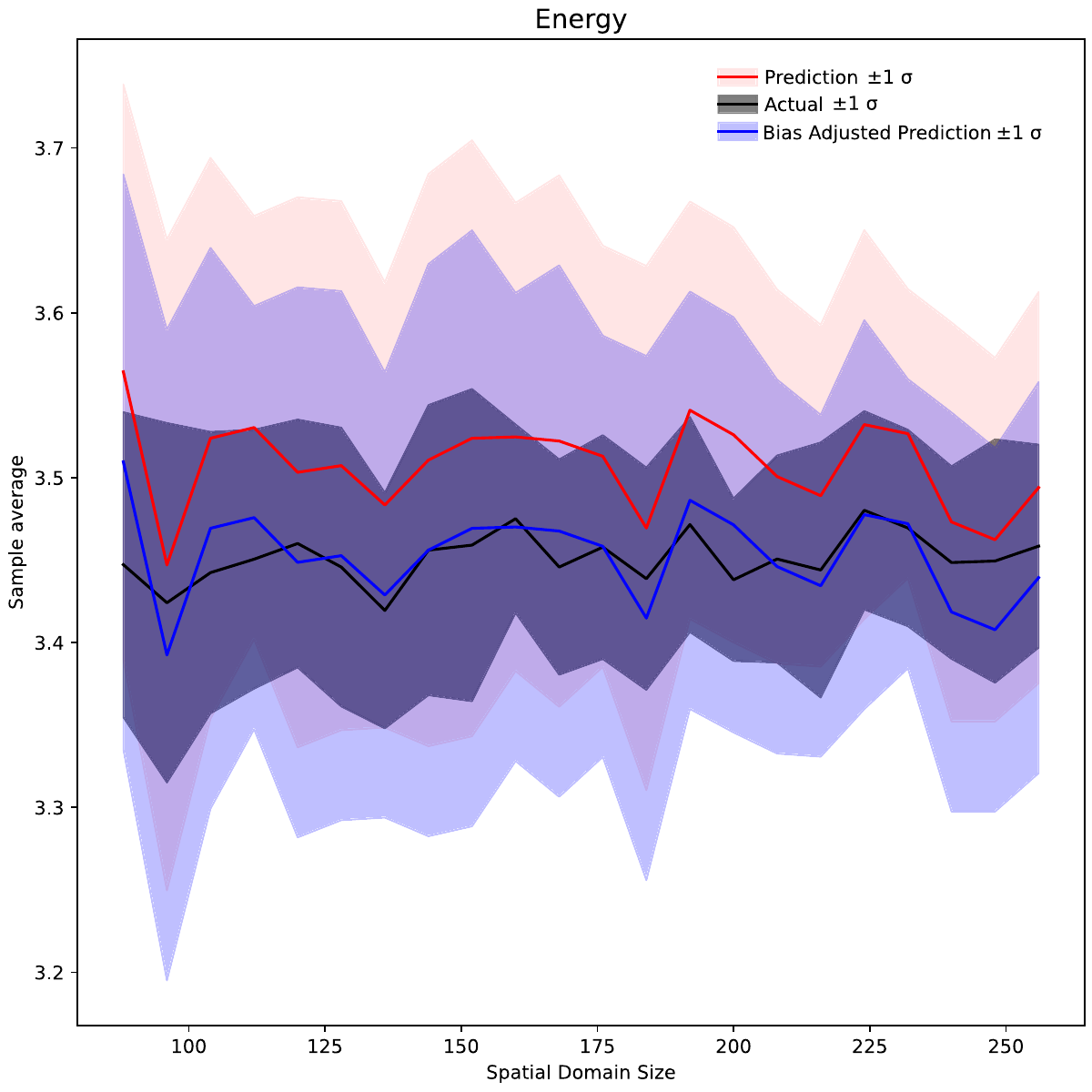}
        \caption{Energy}
    \end{subfigure}
    \hfill
    \begin{subfigure}{0.48\textwidth}
        \centering
        \includegraphics[height=6cm]{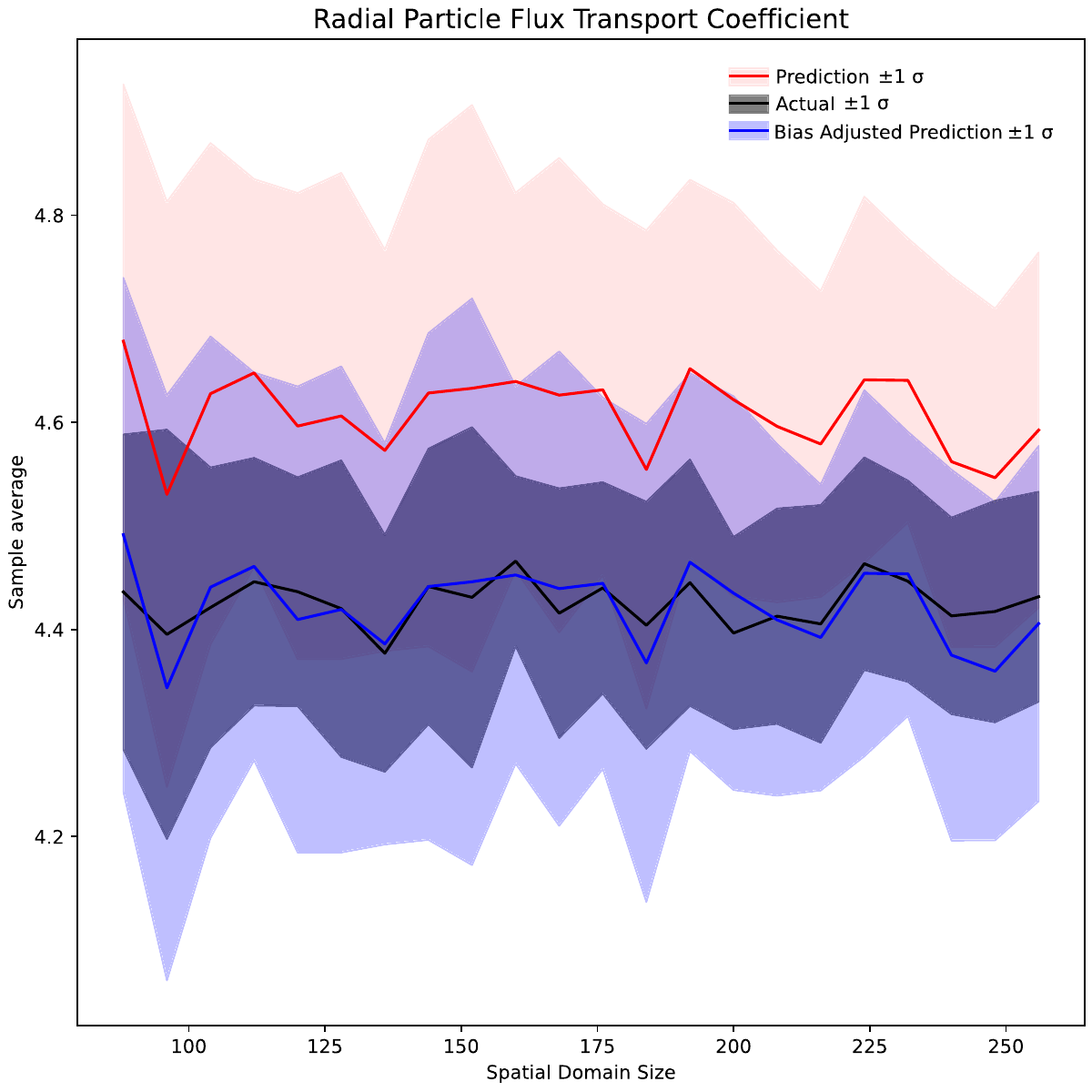}
        \caption{Anomalous Diffusion}
    \end{subfigure}

    \vspace{0.3cm}
    
    \begin{subfigure}{0.48\textwidth}
        \centering
        \includegraphics[height=6cm]{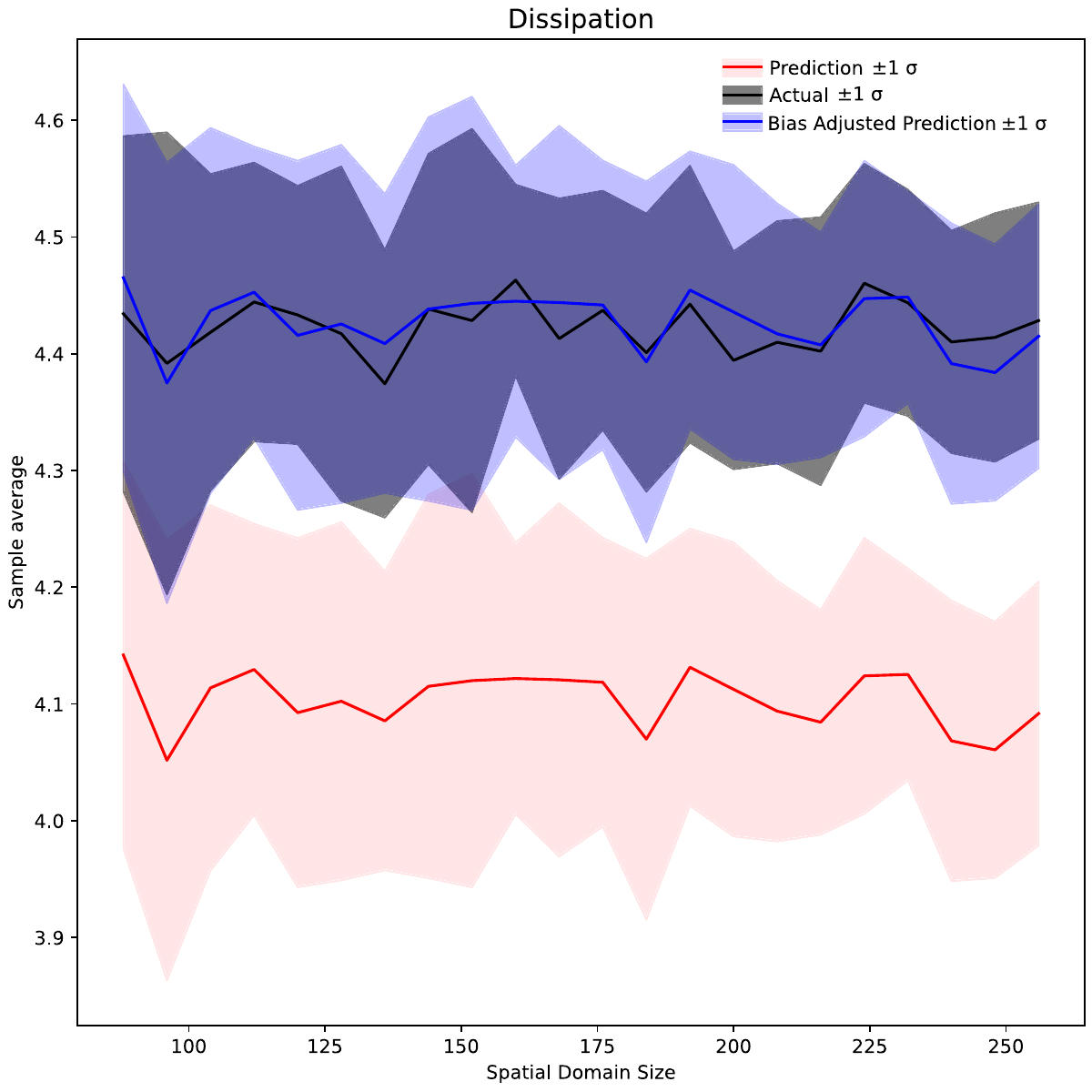}
        \caption{Dissipation}
    \end{subfigure}
    \hfill
    \begin{subfigure}{0.48\textwidth}
        \centering
        \includegraphics[height=6cm]{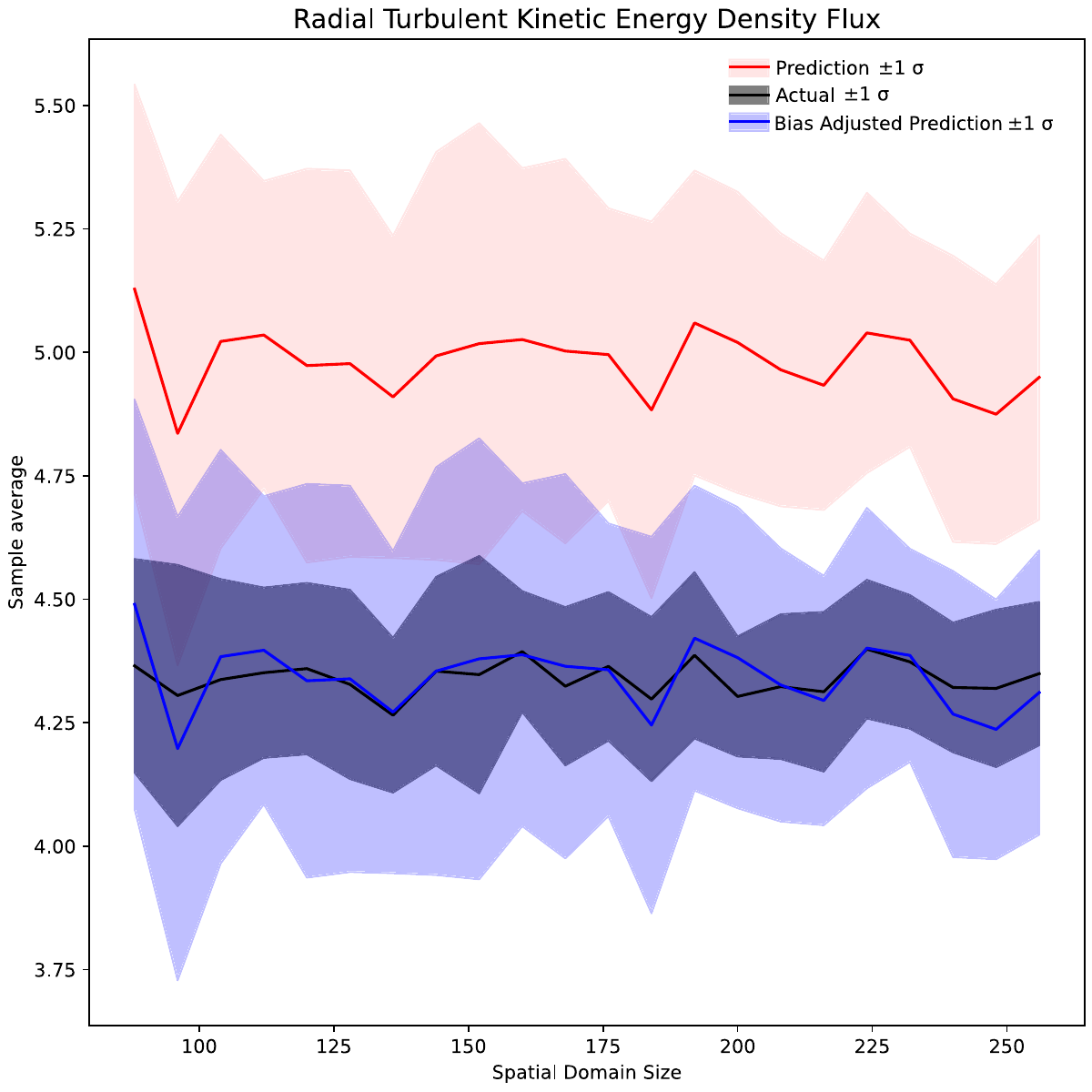}
        \caption{Turbulent Kinetic Energy Density}
    \end{subfigure}

    \vspace{0.2cm}
    
    \caption{\label{DomainScan} Domain scan of average observables that are used to determine the bias offset.}
\end{figure*}

To calculate this bias, we recomputed the weights, and the observables in a range of domain sizes to confirm that the error is largely just a constant offset instead of a systemic error. We then calculated the average difference between the prediction, and the actual average of many runs using data from the $L=144$ domain size. This domain size was chosen as the approximate center of our range of considered domain sizes. Adjusting the red lines in FIGs. \ref{fig:avg_energy_128}-\ref{fig:avg_turbkin_128} by the resulting $L=144$ value yields the depicted blue lines. Confirmation of our hypothesis that the error is largely caused by a (relatively) constant offset can be seen in FIG. \ref{DomainScan}. Where bias adjustment using the $L=144$ value results in consistent agreement across the considered set of domain sizes.

\section{Conclusions and Outlook}
In the preceding sections, we have developed a chaotic field theory description of plasma turbulence, using the Kuramoto-Sivashinsky model of the Trapped Ion Mode as an example. Through this framework, we elucidated the principles of multiperiodic orbit theory, demonstrating how doubly periodic spatiotemporal patterns underpin chaotic motion in spatially extended systems. Utilizing the \textit{Orbithunter} code we identified the fundamental spatiotemporal patterns that characterize a TIM and serve as the building blocks for its dynamics. The importance of each fundamental orbit was then determined through explicitly identifying and tracking how much spacetime each accounted for in a large sample size of arbitrary solutions to the Kuramoto-Sivashinsky equation. This directly translated to providing the weights with which we determined accurate approximations of average potential fluctuation intensity, scaled anomalous diffusion, and turbulent energy flux. 

Focusing on the Trapped Ion Mode allowed for a model choice that was confined to a single field, and 1+1 dimensions. This, in turn, meant that all novel calculations were easy to visualize. Every solution is a 2D picture made up of a mosaic of smaller, simpler 2D pictures. The simplest of which can be cataloged into a finite "alphabet" of fundamental orbits, which form the basis of all turbulence in the system. The importance of each member of the alphabet in reconstructing turbulence can be gleaned by simply observing how frequently each shows up in arbitrary, time-integrated solutions. These frequencies can then be directly translated into weights with which average physical observables may be calculated. Being able to build turbulence out of simple patterns, and then use the same patterns to calculate quantities of interest, is the primary strength of this theory. By breaking the complex behavior into its constituent parts, we are able to make predictions regardless of any specific turbulent dynamics. 

In this work, the building blocks we found are valid for constructing a vast majority of the motion that can be generated by a TIM constrained, of course, by the chosen model. Therefore, the physical averages we calculated are inherent dynamical properties of this type of turbulence--independent of any initial condition. These averages, despite having only a small number of fundamental patterns as support, were in remarkably good agreement with arbitrary, time-integrated TIM turbulence snapshots. This lends further support to the assertion that our alphabet of patterns is sufficient to characterize a TIM, and suggests potential applications within plasma physics.      

Clipping, gluing, and shadowing form the basis of a theory of coherent structures as the building blocks of plasma turbulence. Though applied to a radically reduced model of trapped particle turbulence, there is no part of the theory which breaks when applied to more relevant dynamics. All that is lost is the ease with which the mechanics of the theory can be visualized.  

Reformulating plasma turbulence on the back of fundamental structures and periodic orbit theory presents many challenges. Though this work shows that they not insurmountable, several remain unaddressed. The $L^2$ moving window shadowing calculation, although easy to visualize and understand, neglects the interplay between the shape, and the magnitude of possible shadowing events. This means that patterns in an arbitrary flow could look familiar (eg. a recognizable wave-wave interaction), but the surrounding flow-field could force the magnitude of the potential fluctuation to be unrecognizable to our $L^2$ filter. This could possibly be addressed by modifying out cost function to have separate penalties for disagreement in shape, and magnitude. However, the additional free parameters further obscure the underlying physics, and make confirmation bias more likely. Another option is to move towards a topological metric. However, this introduces its own difficulties. The fundamental periodic orbits are doubly-periodic, whereas the underlying field with which the statistics are determined is not. This vastly complicates topological treatments, and leads to a significant amount of incorrect categorization.   

Despite the challenges of numerically determining shadowing weights, and the inherent approximation choosing a method requires, we have demonstrated that it is possible to determine "typical" values of physical observables with only a small library of simple patterns. Therefore, with the success demonstrated here, it is both warranted and justified to further develop the theory for use on more complex models.

\begin{acknowledgments}
The authors wish to thank Seth Dorfman and Christopher Holland for invaluable discussion on bringing periodic orbit theory to plasma physics, and Cole Stephens for discussion on the physics behind the Trapped Ion Mode. In addition, we would like to thank Predrag Cvitanović for his ongoing support, advice, and insight on our work. This material is based upon work supported by the U.S. Department of Energy, Office of Science, Office of Fusion Energy Sciences, under Award DE-FG02-05ER54809
\end{acknowledgments}

\section*{Data Availability Statement}
The set of fundamental orbits are the most important components to regenerate the results of this paper.
These can be found in a hdf5 file stored in the github repository \cite{orbithunter2025}. They are discretized such that the spatial and temporal grid sizes are as close to 0.5 as possible (in dimensionless units). The integrated trajectories used to generate the shadowing calculations are determined by setting a random seed before generating an initial condition to ensure that results are reproducible. For the n-th sample at a given spatial domain size, the random seeds follow the pattern of $n + 1000 * \text{int}(L)$ .

\appendix

\section{Explicit Form of the Jacobian}
Although obtained through standard computation, the matrix form of the Jacobian \eqref{jacobians} is not immediately obvious. With the discretized form of \eqref{MattKSE} given by \eqref{pseudospecKSE} and \eqref{co-moving}, the first component of the Jacobian is given by

\begin{equation}
    \begin{split}
    \mathbf{J}_{\tilde{u}}\equiv\frac{\partial f}{\partial \tilde{u}}=\textbf{M}[\partial_t]+\textbf{M}[\partial^2_{x}]+\textbf{M}[\partial^4_{x}]+\\
    \textbf{M}[\mathcal{F}_t\partial_x\mathcal{F}_x]\text{diag}[\mathcal{F}^{-1}(\tilde{u})]\mathbf{M}[\mathcal{F}^{-1}]\\
    (\mathbf{J}_S)_{\tilde{u}}\equiv\frac{\partial f_S}{\partial \tilde{u}}=\mathbf{J}_{\tilde{u}}+\frac{S}{T}\mathbf{M}[\partial_x]\\
    \end{split}
\end{equation}
Where $\text{diag}[\mathcal{F}^{-1}(\tilde{u})]$ is a diagonal matrix with $u$ as its elements. Similarly, the first component of the adjoint Jacobian is given by linearizing \eqref{adjointevol}:
\begin{equation}
    \begin{split}
    (\mathbf{J})_{\tilde{u}}^T=-\textbf{M}[\partial_t]+\textbf{M}[\partial^2_{x}]+\textbf{M}[\partial^4_{x}]+\\
        -\mathbf{M}[\mathcal{F}]\text{diag}\left(\mathcal{F}^{-1}(\tilde{u})\right)\mathbf{M}\left[\mathcal{F}^{-1}_t\partial_x\mathcal{F}^{-1}\right]\\
        (\mathbf{J}_{\tilde{u}})^T_S=(\mathbf{J})_{\tilde{u}}^T+\frac{S}{T}\mathbf{M}[\partial_x]
    \end{split}
\end{equation}
Obtaining the other components of \eqref{jacobians} requires the definition of the derivative terms in Fourier basis:
\begin{equation}
\begin{split}
\partial_x \tilde{u} &=
\begin{bmatrix}
q_k c_{jk} & -q_k a_{jk} \\
q_k d_{jk} & -q_k b_{jk}
\end{bmatrix}, \quad
\partial^2_x \tilde{u} =
\begin{bmatrix}
- q_k^2 a_{jk} & - q_k^2 c_{jk} \\
- q_k^2 b_{jk} & - q_k^2 d_{jk}
\end{bmatrix}, \\
\partial^4_x \tilde{u} &=
\begin{bmatrix}
q_k^4 a_{jk} & q_k^4 c_{jk} \\
q_k^4 b_{jk} & q_k^4 d_{jk}
\end{bmatrix}, \quad\hspace{7pt}
\partial_t \tilde{u} =
\begin{bmatrix}
- \omega_j b_{jk} & - \omega_j d_{jk} \\
\omega_j a_{jk} & \omega_j c_{jk}
\end{bmatrix}.
\end{split}
\end{equation}
Where 
\begin{equation}
    \begin{split}
        q_k=-\frac{2\pi k}{L},\quad k\in 1,...\frac{M}{2}-1\\
        \omega_j=-\frac{2\pi j}{T},\quad j\in 1,...\frac{N}{2}-1
    \end{split}
\end{equation}
And $a_{jk},b_{jk},c_{jk},d_{jk}$ are Fourier coefficients. Noting that the $L$ and $T$ dependence is stored in $q_k$ and $\omega_j$ the remaining components of \eqref{jacobians} are given by
\begin{equation}
    \begin{split}
        \frac{\partial f}{\partial T}=-\frac{1}{T}\mathbf{M}[\partial_t]\tilde{u}\\
    \frac{\partial f_S}{\partial T}=\frac{\partial f}{\partial T}-\frac{1}{T}\left(-\frac{S}{T}\mathbf{M}[\partial_x]\tilde{u}\right)\\
    \frac{\partial f}{\partial L}=\left(-\frac{2}{L}\partial^2_{x}-\frac{4}{L}\partial^4_{x}\right)\tilde{u}-\\
    \frac{1}{2L}\mathcal{F}_t\partial_x\mathcal{F}_x\left(\mathcal{F}^{-1}(\tilde{u})\cdot\mathcal{F}^{-1}(\tilde{u})\right)\\
    \frac{\partial f_S}{\partial L}=\frac{\partial f}{\partial L}-\frac{1}{L}\left(-\frac{S}{T}\partial_x\tilde{u}\right)\\
    \frac{\partial f}{\partial S}=-\frac{1}{T}\partial_x\tilde{u}\\
    \end{split}
\end{equation}

\section{Theoretical Underpinnings}
While approached from a data analysis standpoint, the shadowing, and averaging calculations conducted in this paper are based soundly upon the dynamical systems theory laid out in a recent paper by Cvitanović and Liang \cite{Cvitanovic2025chaotic}. The expectation values calculated here (for example \eqref{kinfluxexpec}) are built out of a weighted sum of "Birkhoff" sums of multiperiodic orbits (MPOs) $u_{mpo}$:
\begin{equation}\label{birkavg}
\begin{split}
    A_p[u_{mpo}]=\frac{1}{T_pL_p}\int_0^{L_p}dx\int_0^{T_p}dt(a[u_{mpo}(x,t)])\\
    a_p[u_{mpo}]=\frac{1}{T_pL_p}A_p[u_{mpo}]
\end{split}
\end{equation}
Where $a$ is the observable of interest, $T_p$ is the temporal period, and $L_p$ is the spatial period. Moving forward $T_pL_p=V_p$ will denote the spacetime volume. $A_p$ is the Birkhoff sum, and $a_p$ the Birkhoff average. For $u_{mpo}$ fixed in space or time, there is only one integral, and only one factor in spacetime volume. Due to the density of MPOs in solution space, they can be used as support for the generating function of partition sums for solution volume $V_{c}$:
\begin{equation}\label{partitiongen}
    \begin{split}
        Z[\beta,z]=\sum_cz^{V_{c}}\int_{\mathcal{M}_c}du_{c}\delta\left(F[u]\right)\exp{(V_{c}\vec{\beta}\cdot \vec{a}_p[u_c])}\\
        =\sum_c\frac{z^{V_{c}}}{|\text{Det}\mathcal{J}_{c}|}\exp{(V_{c}\vec{\beta}\cdot \vec{a}_p[u_{mpo}])}
    \end{split}
\end{equation}

Where $F[u]=0$ is the governing field equation \eqref{MattKSE}, the sum over $c$ denotes summing over all periodic states, and $\mathcal{M}_c$ is the infinitesimal neighborhood around $u_{mpo}$. $|\text{Det}\mathcal{J}_{c}|^{-1}$ is the spacetime weight of each multiperiodic orbit, it is the functional determinant of the variation of the governing field equation. The coefficient of a given $z^{V_{\mathbb{A}}}$ in \eqref{partitiongen} is the \textit{primitive cell partition sum} for a tiling of infinite spacetime of primitive cell volume $V_{\mathbb{A}}$. As $V_{\mathbb{A}}\rightarrow\infty$ the primitive cell partition sum, $Z_{\mathbb{A}}[\beta]$,approaches the full infinite spacetime partition sum. Even as the considered primitive volume grows without bound, each $Z_{\mathbb{A}}[\beta]$ is still bounded in the following way
$$Z_{\mathbb{A}}[\beta]\leq e^{V_{\mathbb{A}}(\beta\cdot a_{max}-\lambda_{min}+h_{max})}$$
Where $\lambda_{min}$ encodes the exponentially decreasing spacetime weight, and $h_{max}$ accounts for the exponentially growing number of solutions allowed by larger, and larger spacetime volumes. Therefore, the the generating function \eqref{partitiongen} converges for sufficiently small $z$. Looking at the structure for the primitive cell partition sum, it is possible to extract the appropriate Birkhoff average by taking the logarithmic derivative with respect to $\beta$:
\begin{equation}
    \br{a_{V_{\mathbb{A}}}}=\frac{\partial_{\beta}\ln Z_{\mathbb{A}}\vert_{\beta=0}}{V_{\mathbb{A}}}\equiv\partial_{\beta}W_{\mathbb{A}}\vert_{\beta=0}
\end{equation}
Asymptotically, 
$$W[\beta]=\lim_{V_{\mathbb{A}}\rightarrow\infty}\frac{1}{V_{\mathbb{A}}}\ln Z_{\mathbb{A}}[\beta]$$
This sets the maximum "sufficiently small $z$" that allows convergence of \eqref{partitiongen} to $z_{max}[\beta]=\exp{(-W[\beta])}$, so for $\beta=0$ and$z=z(0)$, $-(dz/d\beta)/z=\br{a}$, the \textit{expectation value}. In other words, $z_{max}[\beta]$ is the first pole of \eqref{partitiongen}. Determining $z_{max}[\beta]$ can be turned into a root-finding problem by defining a "zeta function"

\begin{equation}\label{zeta}
    Z[\beta,z]=-z\partial_z(\ln1/\zeta[\beta,z])
\end{equation}
The construction in \eqref{zeta} leads directly to the first root of $1/\zeta$ being equal to the first pole of $Z[\beta,z]$. Expectation values can now be obtained through the zeta function by taking the root condition, and taking a derivative with respect to $\beta$

\begin{equation}
\begin{split}
    0=1/\zeta[\beta,z_{max}(\beta)]\rightarrow 0=\frac{d}{d\beta}1/\zeta\\=\partial_{\beta}1/\zeta+\frac{dz}{d\beta}\partial_z1/\zeta
\end{split}
\end{equation}

The expectation value is then given as
\begin{equation}\label{obseravg}
    \br{a}=-\frac{1}{z}\frac{dz}{d\beta}\bigg\vert_{\beta=0,z=z(0)}=\frac{\partial_{\beta}1/\zeta}{z\partial_z1/\zeta}\bigg\vert_{\beta=0,z=z(0)}\equiv\frac{\br{Va}_{\zeta}}{\br{V}_{\zeta}}
\end{equation}
Intuitively, \eqref{obseravg} is the "typical" Birkhoff average of observable $a$ for the system. Taking the asymptotic sum of observable values over the MPO support and then dividing by an asymptotic "typical" spacetime volume.

Derivatives of the zeta function are infinite polynomials in $|\text{Det}\mathcal{J}_{c}|^{-1}$. Here, $\br{Va}_{\zeta}$ and $\br{V}_{\zeta}$ were truncated after only considering contributions from the fundamental periodic orbits. These contributions were found by manually sampling the covering of the strange attractor of \eqref{MattKSE}. This computation is equivalent to the determination of $|\text{Det}\mathcal{J}_{c}|^{-1}$. 
\providecommand{\noopsort}[1]{}\providecommand{\singleletter}[1]{#1}%

\end{document}